\documentclass[twocolumn,showpacs,preprintnumbers,amsmath,amssymb%,superscriptaddress
 aps,
 prb,
 lengthcheck,%
]{revtex4-1}

\usepackage{amssymb,color}
\usepackage{graphicx}
\usepackage{amsmath}
\usepackage{hyperref}

\begin{document}

\title{Hanbury Brown and Twiss noise correlations in a \\
topological superconductor beam splitter}

\author{T. Jonckheere,$^1$ J. Rech,$^1$ A. Zazunov,$^2$ R. Egger$^2$ and T. Martin$^1$}
\affiliation{$^1$Aix Marseille Univ, Universit\'e de Toulon, CNRS, CPT, Marseille, France}
\affiliation{$^2$Institut f\"ur Theoretische Physik, Heinrich Heine Universit\"at, D-40225 D\"usseldorf, Germany}

\date{\today}

\begin{abstract}
 We study Hanbury-Brown and Twiss current cross-correlations in a three-terminal junction where a central topological superconductor (TS) nanowire,
 bearing Majorana bound states at its ends, is connected to two normal leads. 
 Relying on a non-perturbative Green function formalism, our calculations allow us to provide analytical expressions for the currents and their correlations at subgap voltages, while also giving exact numerical results valid for arbitrary external bias. 
 We show that when the normal leads are biased at voltages $V_1$ and $V_2$ smaller than the gap, the sign of the current cross-correlations
  is given by $-\mbox{sgn}(V_1 \, V_2)$. In particular, this leads to positive cross-correlations for opposite voltages, a behavior in stark contrast
with the one of a standard superconductor, which provides a direct evidence of the presence of the Majorana zero-mode at the edge of the TS.
We further extend our results, varying the length of the TS (leading to an overlap of the Majorana bound states) as well as its chemical potential (driving it away from half-filling), generalizing the boundary TS Green function to those cases.
 In the case of opposite bias voltages, $\mbox{sgn}(V_1 \, V_2)=-1$,
 driving the TS wire through the topological transition leads to a sign change of the current cross-correlations, providing yet another signature of the physics of the Majorana bound state.

\end{abstract}

\pacs{73.23.-b, 72.70.+m, 74.45.+c}
\maketitle
\section{Introduction}

In the last two decades, Majorana fermions,\cite{majorana_teoria_2008} concepts which initially were the strict property of particle physics,
found some correspondence in condensed matter physics settings. Instead of  looking whether an elementary particle, such as the neutrino,
qualifies as a Majorana fermion, nanoscience physicists are now wondering whether a complex many body electronic system with collective
excitations could bear such strange objects: a fermion whose annihilation operator is (sometimes trivially) related to its creation counterpart.
Indeed, Kitaev\cite{kitaev_unpaired_2000} showed that a one dimensional wire with tight-binding interactions  and  $p$-wave pairing exhibits Majorana
fermions at its boundaries. Recent reviews give a broad summary of this work and its consequences in condensed matter physics.
\cite{alice_new_2012,leijnse_introduction_2012,beenakker_search_2013} There has been an ongoing effort to study experimentally whether this toy model has some correspondence in physical systems. Among strong candidates are nanowires with Rashba and Zeeman coupling put in proximity to
a BCS superconductor,\cite{lutchyn_majorana_2010,oreg_helical_2010,alicea_non_2011,mourik_signatures_2012,das_zero_2012}
and  chains of iron atoms deposited on top
of a lead surface.\cite{nadj-perge_observation_2014}
In Refs.~\onlinecite{mourik_signatures_2012, das_zero_2012}
the signature for the presence of a Majorana fermion at the edge of
 a one dimensional nanowire consists of a zero bias anomaly in the current voltage characteristics. These results call for the exploration
 of more involved transport settings and geometries, where
 the behavior of the Majorana fermion can be fully investigated and characterized. Among these settings,
 multi-terminal hybrid devices offer unique perspectives.
\begin{figure}
\centerline{\includegraphics[width=9.cm]{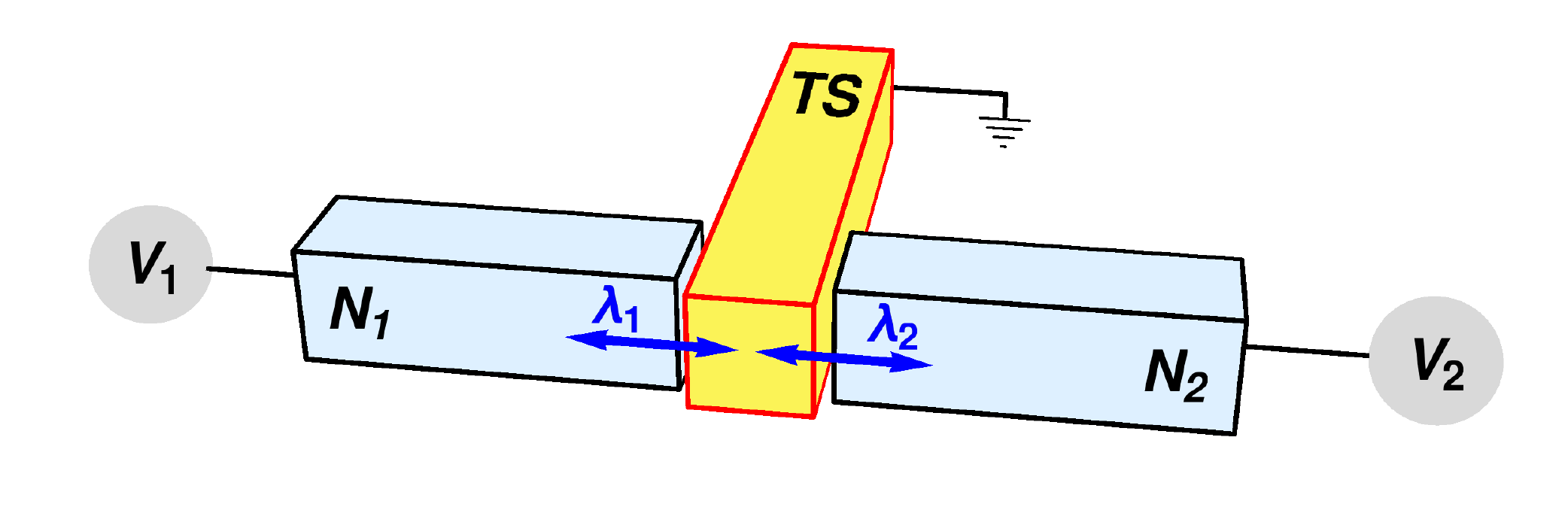}}
\caption{Schematic view of the setup: a grounded TS nanowire
 is tunnel coupled (with hopping amplitudes $\lambda_1$, $\lambda_2$) to two normal-conducting (N$_1$, N$_2$) leads
 which are biased at voltages $V_1$ and $V_2$, respectively.}
\label{fig:setup}
\end{figure}

Multi-terminal devices have often played an important role for exploring the electronic transport properties of 
mesoscopic devices. 
They allow to perform experiments in close analogy with quantum optics scenarios: in the Hanbury-Brown and Twiss \cite{hanburybrown_test_1956} (HBT) experiment for instance, photons impinging on a half-silvered mirror are either transmitted or reflected, and the crossed correlations of intensities from these two outputs are measured, yielding a positive signal due to the bunching of photons emitted from a thermal source. 
Transposed to condensed matter setups, the sign of the HBT cross-correlations reveals meaningful information concerning the physics at play.
For a DC biased three-terminal normal conductor, the electronic analog of the HBT experiment was studied theoretically and experimentally two decades ago. 
\cite{buttiker_scattering_1990,
martin_wave-packet_1992,buttiker_scattering_1992,henny_fermionic_1999,oliver_hanbury_1999}
This was analyzed in terms of current-current crossed correlations (noise): fermion antibunching (resulting from Pauli principle) leads to a negative HBT noise signal.
In the context of conventional BCS superconductivity, three-terminal devices consisting of a superconductor connected to two normal leads were also investigated.
In such devices, a Cooper pair can be transferred as a whole in one of the normal leads (via Andreev Reflection or AR),
or it can be split into its two constituent electrons in opposite leads (via Crossed Andreev Reflection or CAR).
The HBT noise correlations can thus be negative or positive depending on whether AR or CAR
is dominant. \cite{martin_wave_1996,torres_positive_1999}
By adding appropriate filters (in energy or spin) to the device in order to rule out AR in each lead, positive noise crossed correlations,
due solely to CAR processes, can be guaranteed. \cite{lesovik_electronic_2001,chevallier_current_2011}
Experimental evidence for Cooper pair splitting in BCS superconductors has been found both in non-local
current measurements \cite{hofstetter_cooper_2009, herrmann_carbon_2010,schindele_near_2012} as well as in noise correlation
measurements\cite{das_high_2012} with a device analogous to what was proposed
in Refs.~\onlinecite{recher_andreev_2001,lesovik_electronic_2001}.                             

In this work, we study HBT noise correlations for a pair splitter, where a TS nanowire,
rather than a standard BCS
superconductor, is connected to two biased normal leads. A schematic
view of the setup is shown in Fig.~\ref{fig:setup}. Because of the presence
of the Majorana zero-mode for the TS, the AR and CAR processes are strongly affected.
Studying the current and the HBT crossed-correlations in this setup, one can expect to explore properties and manifestations of the Majorana bound state.
This setup was previously studied
using scattering theory and tight-binding numerical calculations.\cite{haim_signatures_2015,haim_current_2015,valentini_finite_2016}
However, these works concentrated on the specific case of equal voltages for the normal
leads and focused on voltages below the gap.
Our goal is to provide a complete description of the system, exploring the behavior
of the current and current correlations, for arbitrary values of the voltages, thus capturing both the effect of the Majorana bound state and the high-energy quasiparticles.
Below the gap, we confirm that a TS beam splitter has negative HBT correlations when the two normal leads have equal voltages.\cite{haim_current_2015,valentini_finite_2016}
More importantly, we also predict a reversal of the sign of the current correlations when voltages are changed from equal to opposite, a feature that is directly related to the properties of the Majorana bound state at the end of the TS nanowire.

We perform the calculations within a phenomenological tunnel Hamiltonian approach,
using the boundary  Keldysh Green function of a semi-infinite TS. Solving non-perturbatively the Dyson equation, this allows us to obtain analytical
formulas for the current and current correlations. This approach was
introduced in Ref.~\onlinecite{zazunov_low_2016}, and can be used to treat a system
composed of an arbitrary number of leads. The boundary Keldysh Green function of the TS
nanowire encapsulates all the properties of the TS boundary. The corresponding density of states shows a zero-energy peak, associated with the Majorana zero-mode, and a non-zero density of states above the proximity-induced gap without BCS singularity.

Moreover, we can further emphasize the role played by the Majorana bound state by generalizing the TS boundary Green function to the case of a finite length nanowire, or one with an arbitrary chemical potential (adjustable doping). For a TS of finite length $L$, the two Majorana bound states localized at the opposite ends of the nanowire can overlap, leading to a vanishing of its effect for small enough $L$, which we confirm by looking at the $L$-dependence of the HBT correlations. Also, tuning the chemical potential of the TS allows us to drive the transition from a topological regime to a trivial (non-topological) one, which is also manifest in the current correlations.

The structure of the paper is as follow. In Sec.~\ref{sec:model} we introduce the Hamiltonian model and detail the formulas for the transport properties.
Sec.~\ref{sec:results} is devoted to our main results. First, analytical expressions and numerical results for the current and
differential conductances are presented and discussed. These quantities can readily be measured in experimental setups. We then provide explicit expressions for the current correlations in the subgap regime, along with numerical results at any bias.
A detailed qualitative discussion of the noise behavior, in relation to the particular properties of the Majorana bound state is also given.
Sec.~\ref{sec:beyond} focuses on finite size effects as well as the impact of tuning the chemical potential of the nanowire, driving it away from half-filling. In App.~\ref{app:Green_mu}, the derivation of the boundary Green function for a TS nanowire with a finite bandwidth and arbitrary chemical potential is presented.
General analytical expressions for the current and current correlations are
provided in App.~\ref{app:formulas}.
Finally, App.~\ref{app:Smatrix} discusses subtleties of the microscopic tunneling model,
and establishes its equivalence with the scattering matrix formalism for subgap voltages.

\section{Model and Formalism} \label{sec:model}

We consider a three-terminal device in a T-shaped geometry, as illustrated in Fig.~\ref{fig:setup},
where the end of a topological superconductor (TS) nanowire is contacted by two normal-conducting (N) leads.
%This setup has also been studied in other theoretical works, for instance, see recent Refs.~[von Oppen, Fazio].
In a general case, two different voltages $V_{1,2}$ are applied across the N-TS contacts,
while the TS wire is assumed to be grounded. The full Hamiltonian is given by
\begin{equation}
H = H_{TS} + H_N + H_t ,
\end{equation}
where the first two terms describe the TS and two N-leads,
respectively, and $H_t$ is a tunneling Hamiltonian connecting all three leads to each other (see below for details).
We model the TS wire as a semi-infinite 1D spinless $p$-wave superconductor,
corresponding to the continuum version of a Kitaev chain\cite{kitaev_unpaired_2000,alice_new_2012}
in the wide-band limit.
The Hamiltonian of the TS wire located at $x>0$ reads
\begin{equation}
H_{TS} = \int_0^{\infty} \!\!\!\! dx \; \Psi^{\dagger}_{TS}(x) \left( -i v_F \partial_x \sigma_z
 + \Delta \sigma_y \right) \Psi_{TS}(x) ,
\end{equation}
where $\Delta$ is a proximity-induced pairing gap, assumed to be real,
the Nambu spinor $\Psi_{TS}(x) = (c_r,c^{\dagger}_l)^T$ combines right- and left-moving fermions,
with annihilation field operators $c_r(x)$ and $c_l(x)$, respectively, $v_F$ is the Fermi velocity and
$\sigma_{x,y,z}$ are the Pauli matrices in Nambu space. In what follows, we use units with $k_B = v_F = \hbar = 1$.

In this work, following the approach of Ref.~\onlinecite{zazunov_low_2016},
we formulate the transport problem in terms of boundary Keldysh Green functions (GFs)
describing the leads which are coupled together by tunneling processes.
For such a noninteracting setup with point-like tunneling contacts,
the exact boundary GFs can be obtained by solving the Dyson equation to all orders
in the tunnel couplings. Below we briefly review the boundary GF approach
(see Ref.~\onlinecite{zazunov_low_2016} for details) and
summarize relevant formulas needed for the calculation of transport observables in
the three-terminal N-TS-N junction.

The boundary Keldysh GF at $x=0$ for the TS wire is defined as follows:
\begin{equation}
\check{g}_{TS}(t - t')= -i \left \langle {\cal T}_C \Psi(t) \Psi^{\dagger}(t') \right \rangle ,
\end{equation}
where the Nambu spinor $\Psi=(c, c^{\dagger})^T$ contains the boundary fermion operator
$c = [c_l + c_r] (x=0)$, and ${\cal T}_C$ denotes Keldysh time ordering. Explicitly, the Fourier transforms of the retarded and advanced
GFs for the uncoupled TS wire in the topologically nontrivial phase derived in Ref. \onlinecite{zazunov_low_2016} are
\begin{align}
g^{R/A}_{TS}(\omega) =&  \frac{\sqrt{\Delta^2 - (\omega \pm i 0^+)^2} \, \sigma_0  + \Delta \sigma_x}
                          {\omega \pm i 0^+} ,
\label{eq:Ginfinite}
\end{align}
where $R/A$ corresponds to $+/-$ and $\sigma_0$ is the unity matrix in Nambu space.
Importantly, this simple expression for the retarded/advanced boundary GF of a TS wire
captures the zero-energy Majorana bound state as well as continuum quasiparticles,
which allows for studying both subgap and above-gap transport on equal footing.
The Keldysh component $g^K_{TS}(\omega)$ is expressed via the retarded and advanced components as
\begin{align}\label{eq:gK}
g^{K}_{TS}(\omega) =& (1-2 n_F(\omega)) \left[ g^{R}_{TS}(\omega) - g^{A}_{TS}(\omega) \right] ,
\end{align}
where $n_F(\omega) = \left( e^{\omega / T} + 1 \right)^{-1}$ is the Fermi function with
temperature $T$. Throughout the paper we use the chemical potential $\mu_{TS}$ of the (grounded) TS wire as a reference energy level
and set $\mu_{TS}=0$.
%Notice that $\mu_{TS}$ should not to be confused with the parameter $\mu$ of the Kitaev chain model
%characterizing the conduction band filling of an uncoupled TS wire.
In App.~\ref{app:Green_mu}, we give a derivation of the boundary Green function for a Kitaev
chain with the finite bandwidth and arbitrary values for the band filling,
while the wide-band expression \eqref{eq:Ginfinite} exhibiting particle-hole symmetry corresponds to half filling.

In the same manner we construct the Keldysh GFs for the normal leads.
Within the wide-band approximation,
taking into account that the N-TS tunnel coupling effectively involves only one spin component in the normal conductor,
the retarded/advanced GF for the normal electrodes follows from Eq.~\eqref{eq:Ginfinite} by putting $\Delta = 0$,
\begin{equation}
g^{R/A}_{N}(\omega) = \mp i \sigma_0 ,
\end{equation}
%and we have silently assumed that the normal density of states is constant near the Fermi level.
Correspondingly, the Keldysh component $g^K_N(\omega)$ is determined via $g^{R/A}_{N}(\omega)$ by a
relation similar to Eq.~\eqref{eq:gK} but with the respective chemical potential $\mu_N$ in the Fermi function
(matrix in Nambu space),
\begin{align}
g^{K}_N(\omega) =& - 2 i \left[ 1-2 n_F(\omega - \mu_N \sigma_z) \right] \sigma_0 .
\end{align}
In the voltage-biased junction, $\mu_N$ is shifted with respect to $\mu_{TS} = 0$ by the dc voltage across the N-TS contact.

In terms of the boundary fermions $c_j$ representing the three leads,
with $j=0$ for the TS wire and $j=1,2$ for the normal-conducting electrodes,
the tunneling Hamiltonian takes the form~\cite{zazunov_low_2016}
\begin{equation}\label{Ht}
H_t = \frac{1}{2} \sum_{j, j'} \Psi^{\dagger}_j W_{j j'} \Psi_{j'} ,
\end{equation}
with $\Psi_{j} = (c_j, c^{\dagger}_j)^T$ the boundary Nambu spinor and
$W = W^\dagger$ is the tunneling matrix in lead and Nambu space.
In lead space, we impose that $W$ has vanishing diagonal elements $W_{j j} = 0$ for all $j$,
while the off-diagonal elements of $W$ are matrices in Nambu space, $W_{j j'} = \lambda_{j j'} \sigma_z$,
with a hopping amplitude $\lambda_{j j'}$.
For our setup when two normal electrodes are connected to the central TS lead,
the only non-vanishing couplings are $\lambda_{0 j} = \lambda_{j 0}^\ast = \lambda_j$ for $j=1,2$.
Without loss of generality, $\lambda_{1,2}$ can always be chosen real, and
in the case of a single tunnel junction
they determine the normal transmission probability of the respective N-TS contact. \cite{zazunov_low_2016}

Once the tunneling matrix $W$ is specified,
the full Keldysh GF $\check G$ of the system follows from the Dyson equation
\begin{equation}
\check{G} = (\check{g}^{-1} - \check{W})^{-1} ,
\label{eq:Dyson}
\end{equation}
with the Keldysh matrix $\check{W} = \mbox{diag}(W,-W)$, and
the ``uncoupled'' Keldysh GF $\check g$ is diagonal in lead space.
From the tunneling Hamiltonian \eqref{Ht}, it is straightforward to get the Heisenberg operator for the current
flowing from lead $j$ to the contact region,
\begin{equation}
\hat{I}_j(t) = i \frac{e}{\hbar} \sum_{j' \neq j} \Psi^{\dagger}_j(t) \sigma_z W_{j j'} \Psi_{j'}(t) ,
\end{equation}
while the dc current $I_j = \langle \hat{I}_j(t) \rangle$ is
 \begin{equation}
 I_j = \frac{1}{2} \frac{e}{\hbar} \int_{-\infty}^{\infty}{d \omega \over 2 \pi} \; \sum_{j' \neq j}\mbox{tr}_N \left[
   \sigma_z W_{j j'} G^K_{j' j}(\omega) \right] ,
   \label{eq:curJfromG}
 \end{equation}
 where tr$_N$ is the trace over Nambu space.
The Keldysh component of the full GF, $G^K$, is given by~\cite{zazunov_low_2016}
\begin{align}
  G^K(\omega) = & G^R(\omega) F(\omega) - F(\omega) G^A(\omega)  \nonumber \\
  & + G^R(\omega) \left[ F(\omega) W - W F(\omega) \right]  G^A(\omega),
\end{align}
where $F_{jk}(\omega) = \delta_{jk} \left[ 1-2 n_F(\omega-\mu_j \sigma_z)\right]$ contains the distribution functions
of the uncoupled leads with the respective chemical potentials $\mu_j$.
 
 Finally, the HBT correlations are readily obtained through the same formalism by computing the zero-frequency cross-correlations of the above-defined current operator. Quite generally, the current correlations at zero frequency are defined as
 \begin{equation}
  S_{j j'} = \int_{-\infty}^{\infty} \!\!\!\!\!  d\tau \;\left \langle
   \delta \hat{I}_j(\tau) \delta\hat{I}_{j'}(0) \right \rangle ,
   \label{eq:SjjfromG}
\end{equation}
with $\delta \hat{I}_{j}(t) = \hat{I}_j(t) - I_j$.
In terms of the full GF, these current correlations are given by~\cite{zazunov_low_2016}
\begin{align}
 S_{j j'} = \int_{-\infty}^{\infty} {d \omega \over 2 \pi} \, \sum_{j_1 \neq j} \sum_{j_2 \neq j'}
  \mbox{tr}_N \bigg\{ \lambda_{j j_1} \Big[ G^{- +}_{j_1 j_2}(\omega) \lambda_{j_2 j'}
     G^{+ -}_{j' j}(\omega)
      \nonumber \\ 
       -G^{- +}_{j_1 j'}(\omega) \lambda_{j' j_2} G^{+ -}_{j_2 j}(\omega)  \Big] \bigg\} ,
\end{align}
where $G^{-s \, s} = (1/2) \left[ G^K + s \left( G^R-G^A \right) \right]$ with $s = \pm$.

\section{Results} \label{sec:results}

As the system consists of three electrodes, with 0-1 and 0-2 couplings only, it is possible to solve explicitly the Dyson equation, Eq.~\eqref{eq:Dyson}, in order to obtain an analytical expression for the full Green function $\check{G}$. 
Using Eqs.~\eqref{eq:curJfromG} and \eqref{eq:SjjfromG}, one can
then derive an explicit form for the average  current $I_j$, and the current correlations $S_{j j'}$. The expressions we give below
depend on the couplings $\lambda_1$, $\lambda_2$ between the TS and the normal electrodes 1 and 2, on the voltages $V_1$ and $V_2$ of the two normal electrodes, and on the temperature through the Fermi function $n_F(x)$.
A natural quantity appearing in these formulas is
$\Lambda = \sqrt{\lambda_1^2+\lambda_2^2}$,
which is related to the total transmission probability $\tau$ between the
TS and the normal leads: $\tau= 4 \Lambda^2/ (1+\Lambda^2)^2$.
For all the results presented in this work, we focus on the case $0<\Lambda <1$,
which covers all possible values of the transmission $\tau \in [0,1]$.
Taking $\Lambda>1$ would give the same value of $\tau$, but for a different realization of the physical system - see the discussion in App.~\ref{app:Smatrix} for more details.

\subsection{Current and differential conductance}

The current flowing through normal lead $j=1,2$ can be written in the simple form:
%facteur (-2) ajoute par-rapport aux expression Mathematica, cf. confusion entre f et nf
\begin{equation}
I_j =\frac{e}{h} \int_{-\infty}^{\infty} d\omega \sum_{k=1,2}
\sum_{s = \pm} s~ n_F(\omega- s ~eV_k) J_{jk}(\omega)
\label{eq:Igeneral}
\end{equation}
where $J_{jj}$ and $J_{j,k \neq j}$ determine, respectively, the local and non-local differential conductances at zero temperature.
%Similar formulas for $I_2$ are of course obtained by exchanging 1 and 2.

 Focusing e.g. on current $I_1$, it can be separated into a ``direct'' contribution
 related to the chemical potential of electrode 1 ($J_{11}$), and a ``non-local'' contribution related to the chemical potential
 of electrode 2 ($J_{12}$). The explicit expressions for
 the differential conductances
 $J_{11}(\omega)$ and $J_{12}(\omega)$ for $|\omega|<\Delta$ are
\begin{align}
 J_{11}(\omega) &= 
 \frac{4 \lambda_1^2 \Lambda^2} {\left(1-\Lambda^4\right)^2 \frac{\omega^2}{\Delta^2}+ 4\Lambda^4}
 - J_{12}(\omega)  , \nonumber \\
  J_{12}(\omega) &=\frac{2 \lambda_1^2 \lambda_2^2 \left(1-\Lambda^4\right)
  \frac{\omega^2}{\Delta^2}} {\left(1-\Lambda^4\right)^2 \frac{\omega^2}{\Delta^2}+4 \Lambda^4},
   \label{eq:J11_J12}
\end{align}
while the expressions for $|\omega| > \Delta$ are given in App.~\ref{app:formulas}.
One can see that in the low-voltage regime, the contribution to $I_1$ from $J_{11}$ is linear in $V$, while the one from $J_{12}$ scales as $\sim V^3$ therefore not contributing to the linear conductance
in the zero-temperature limit. 
This means in particular that in the low voltage limit, 
when coupling to the TS occurs through the Majorana bound 
state only, the current in one normal electrode is not 
influenced by the voltage in the other one. \footnote{In a specific system, depending on the geometry, there might be a trivial direct current, linear in $V$,  if there is a direct coupling between the two normal electrodes for the spin component which is not coupled to the topological superconductor.}

 One can also check that when setting $\lambda_2=0$, Eq.~\eqref{eq:J11_J12} gives back the known formula for a single N-TS junction. While $J_{12}$ trivially vanishes in this case, $J_{11}$ reduces to $1/(1+\omega^2/\Gamma^2)$ for $|\omega|<\Delta$, with $\Gamma = 2 \Delta \Lambda^2/(1- \Lambda^4)$.\cite{zazunov_low_2016}
\begin{figure}
\centerline{\includegraphics[height=4.25cm]{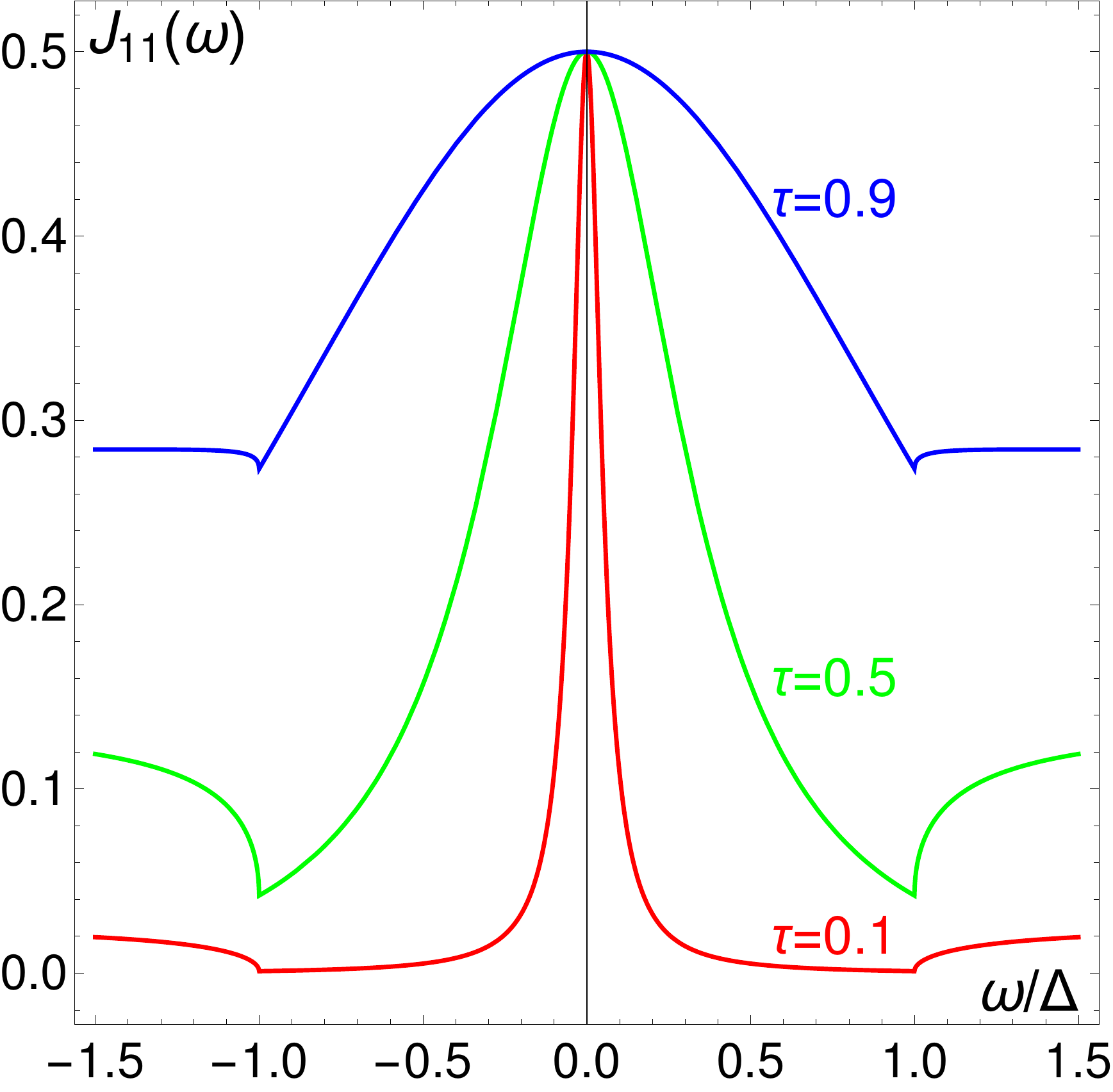} %in ""current_analytics.nb"
\includegraphics[height=4.25cm]{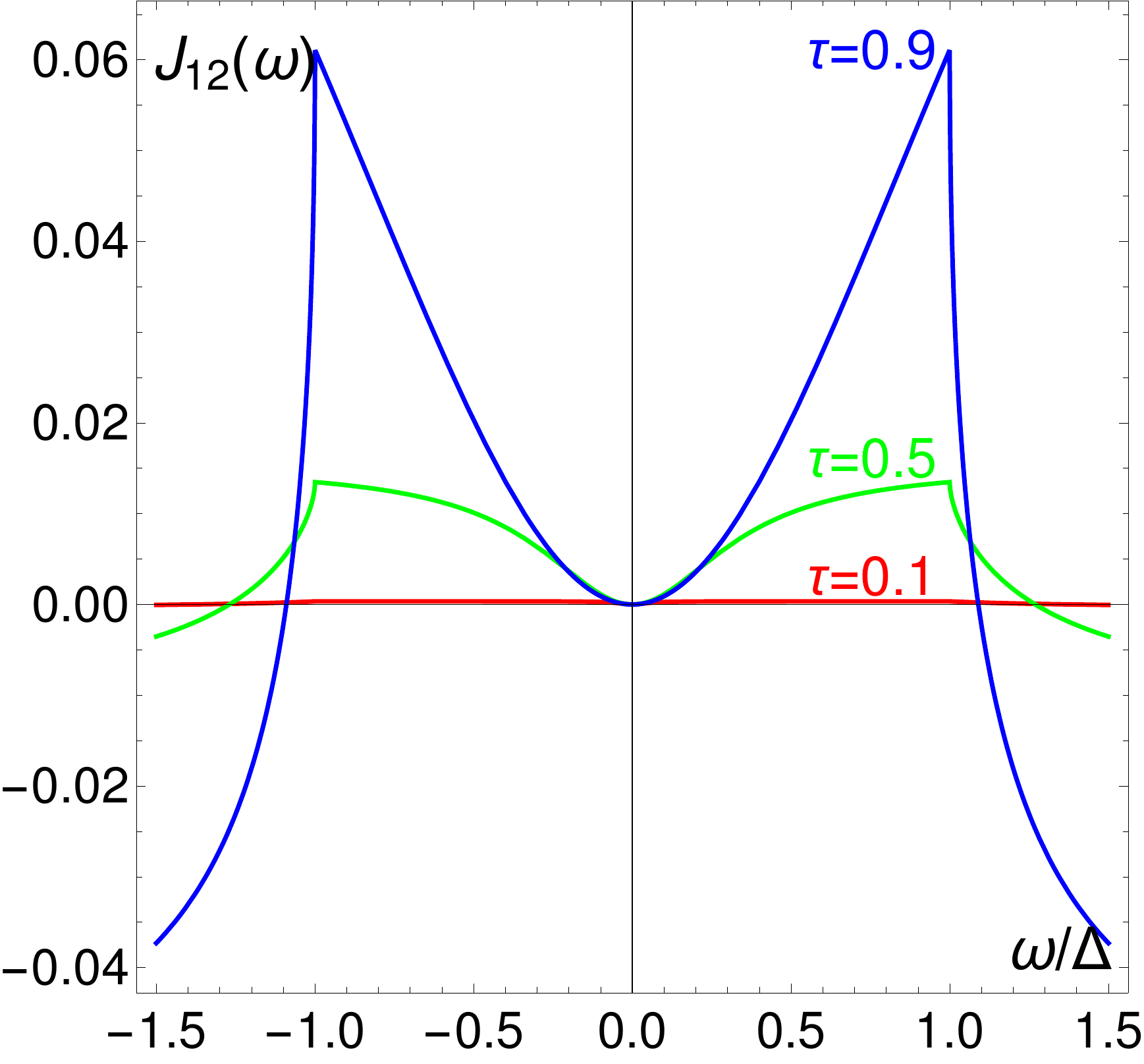}}
\caption{Zero-temperature local $J_{11}(\omega)$ (left panel) and non-local $J_{12}(\omega)$ (right panel) differential conductances
(in units of $2 e^2/h$) vs $\omega/\Delta$ for $\lambda_1=\lambda_2$ and several values of the
total transmission probability $\tau$.}
\label{fig:J11_J12}
\end{figure}

Fig.~\ref{fig:J11_J12} shows the local and non-local differential conductances $J_{11}(\omega)$
and $J_{12}(\omega)$, for three values of the transmission $\tau$. Focusing
for simplicity on the zero-temperature limit, the differential conductance $G_1= dI_1/dV_1$ is given by
\begin{align}
 G_1(V_1) = \frac{2 e^2}{h}   J_{11}(V_1) \simeq
 \frac{2 e^2}{h} \frac{\lambda_1^2/\Lambda^2}{1+V_1^2/\Gamma^2} .
\label{eq:G1}
\end{align}
The factor $\lambda_1^2/\Lambda^2$ reduces to $1/2$ for equal couplings $\lambda_1=\lambda_2$, and is otherwise related to the asymmetry of the
couplings. 
The local differential conductance $J_{11}(\omega)$ has a shape which is similar to the one of a simple N-TS junction, with a peak associated with the Majorana bound state, broadened by the couplings to the normal electrodes. Indeed, the Lorentzian factor $(1+V_1^2/\Gamma^2)^{-1}$ in Eq.~\eqref{eq:G1} is reminiscent of the well-known conductance peak of width $\Gamma$, and height $2 e^2/h$.\cite{law_majorana_2009,wimmer_quantum_2011}
Here, the contribution of this peak is split between the two normal electrodes, resulting in an extra factor $1/2$ in the equal coupling case, so that the zero-voltage differential conductance is simply $e^2/h$ for each electrode.
The non-local differential conductance $J_{12}$ is shown on the right panel of
Fig.~\ref{fig:J11_J12}, for the case of a symmetric junction,
and for three different values of the total transmission $\tau$. 
$J_{12}$ is negligible for very small transmission ($\tau=0.1$).
 For larger transmissions, it starts from 0 at $\omega=0$, and is positive for $|\omega| < \Delta$. $J_{12}$ increases as $\omega$ gets closer to $\Delta$, then abruptly changes sign above the gap, becoming negative for $|\omega| > \Delta$.

\begin{figure}
\centerline{\includegraphics[width=8.5cm]{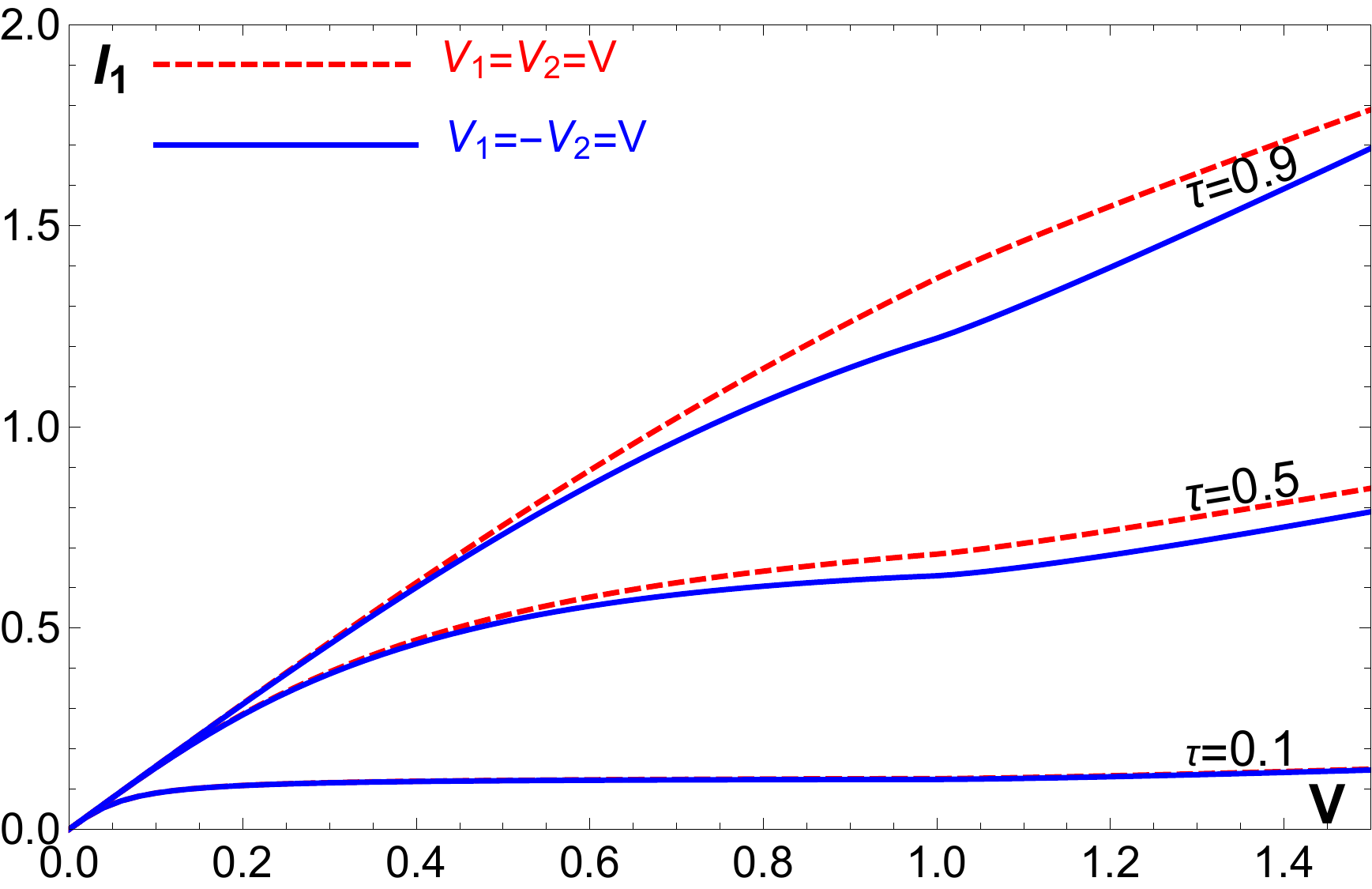}}  %in "current_analytics_bis.nb"
\caption{Current $I_1$ (in units of $e\Delta/h)$ vs voltage $V = V_1$ in the case of equal (red dashed curve) 
and opposite (blue full curve) bias voltages $V_{1,2}$
at zero temperature, $\lambda_1=\lambda_2$ and several values of $\tau$. In all figures, voltages are given
in units of $\Delta$.}
\label{fig:CurV}
\end{figure}

To obtain simple, easily readable formulas for the current,
 it is useful to take the zero temperature limit,
and consider symmetric couplings $\lambda_1 = \lambda_2 = \Lambda/\sqrt{2}$. We then get for the current at voltages below the gap:
\begin{align}
I_1 = &  \frac{e \Gamma}{2h} \left\{ \left( 1+ \sqrt{1 + \frac{\Gamma^2}{\Delta^2}} \right)
\mbox{tan}^{-1}\left(\frac{eV_1}{\Gamma}\right) \right. \nonumber \\
 &  \left. + \left( 1 -  \sqrt{1 + \frac{\Gamma^2}{\Delta^2}} \right)
    \left[ \frac{eV_1 - eV_2}{\Gamma } + \mbox{tan}^{-1}\left(\frac{eV_2}{\Gamma}\right)  \right] \right\}
    \label{eq:I1full}
\end{align}
which for equal voltages $V_1=V_2=V$ reduces to:
\begin{equation}
I_1 = \frac{e}{h} \, \Gamma \, \mbox{tan}^{-1}\left(\frac{eV}{\Gamma}\right)
\end{equation}
Fig.~\ref{fig:CurV} shows numerical results for the current $I_1$ as a function of $V$ for a symmetric junction in the case of equal ($V_1 = V_2 = V$) and opposite voltages ($V_1 = -V_2 = V$).
 Here we focused on three different values of the total transmission between the TS and the normal leads ($\tau = 0.1, 0.5, 0.9$) and 
 for simplicity we considered the case of zero temperature.
The differences in the current $I_1$ between the $V_2=V_1$ and $V_2=-V_1$ cases can
be understood from the properties of the non-local differential conductance $J_{12}$
[see Eq.~\eqref{eq:J11_J12} and the right panel of Fig.~\ref{fig:J11_J12}].
For small transparency ($\tau=0.1$), $J_{12}$ is very small, and the two currents cannot
be distinguished in the figure. For larger transparency, the effect of $J_{12}$ is only noticeable for large enough voltage. As $J_{12}(\omega) > 0$ for $|\omega| < \Delta$,
the current for $V_{2} = V_{1}$ is larger than the one for $V_{2}=-V_{1}$. The difference between the two currents is maximal for $V$ close to $\Delta$, when $J_{12}$ reaches its maximum, before decreasing at higher voltages since $J_{12}(\omega)$ becomes negative at high energy.

For arbitrary subgap voltages, the current $I_2$ is readily obtained from Eq.~\eqref{eq:I1full} upon exchanging $V_1$ and $V_2$. For symmetric couplings $\lambda_1 = \lambda_2$, one can easily convince oneself that $I_2 = I_1$ in the specific case of equal voltages, while in the opposite voltage case, one has $I_2 = -I_1$.

\subsection{Hanbury-Brown and Twiss cross-correlations}

As we did for the current, we derive here analytical expressions for the current auto- and cross-correlations $S_{j j'}$,
and we are particularly interested in the HBT cross-correlations $S_{12}$ between the currents flowing from the central TS to the
two normal leads. As the formulas get rapidly long and cumbersome, we give here the
zero-temperature expression valid for voltages below the gap $\Delta$, and for equal couplings
$\lambda_1=\lambda_2=\lambda$ (thus $\Lambda^2 = \lambda_1^2+\lambda_2^2= 2\lambda^2$).
More general formulas are given in App.~\ref{app:formulas}.
 Introducing as before the broadening $\Gamma =  2 \Delta \Lambda^2/(1- \Lambda^4)$,
 and taking without loss of generality $|V_1| \geq |V_2|$, we have:
\begin{multline}
 S_{12}(V_1,V_2) = \frac{e^2}{h} \frac{\Gamma^2}{4 \Delta^2} \Bigg\{
                  \left( 1 + \frac{1}{2} \frac{\Gamma^2}{\Gamma^2 + (eV_1)^2} \right) |eV_1| \\+
                             \frac{1}{2} \frac{\Gamma^2|eV_2|}{\Gamma^2 + (eV_2)^2}
                             - \frac{3 \Gamma}{2} \mbox{tan}^{-1}\left(\frac{|eV_1|}{\Gamma}\right)
                             - \frac{\Gamma}{2} \mbox{tan}^{-1}\left(\frac{|eV_2|}{\Gamma}\right)
                                                   \\
               - \, \mbox{sgn}(V_1 V_2) \left[
                |eV_2|  \frac{(eV_2)^2 +2 \Gamma^2 + 2 \Delta^2 }{(eV_2)^2 + \Gamma^2}
                -  2 \Gamma \mbox{tan}^{-1}\left(\frac{|eV_2|}{\Gamma}\right) \right]
                 \Bigg\}
\label{eq:S12gen}
\end{multline}

The last term of this expression is proportional to $-\mbox{sgn}(V_1 V_2)$,
with a coefficient which is always positive, independently of the voltages. As it turns out, this term is dominant, and gives the sign of $S_{12}$ for all voltages below the gap.
In the limit of small
voltages $|V_1|, |V_2| \ll \Gamma$, $S_{12}$ becomes
\begin{equation}
 S_{12} \simeq - \frac{1}{2} \frac{e^2}{h}  \, \mbox{sgn}(V_1 V_2) |eV_2|
\end{equation}
Importantly, this means that the HBT cross-correlations are positive when the two voltages
$V_1$ and $V_2$ have opposite signs.

\begin{figure*}
% Thib : figures made in helicity_..._18052015.nb
 \begin{tabular}{cc}
  \includegraphics[width=7.cm]{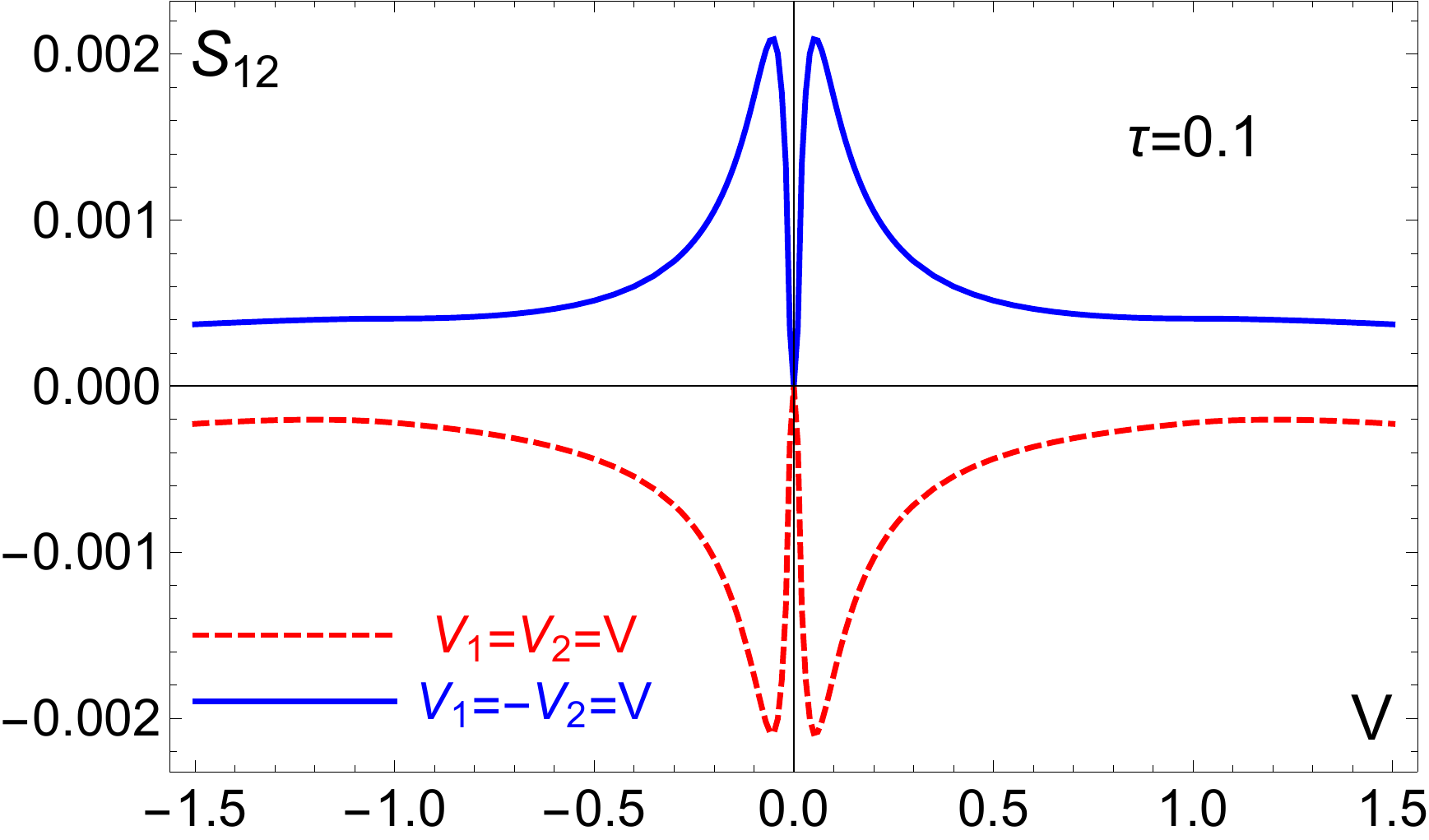} & \includegraphics[width=7.cm]{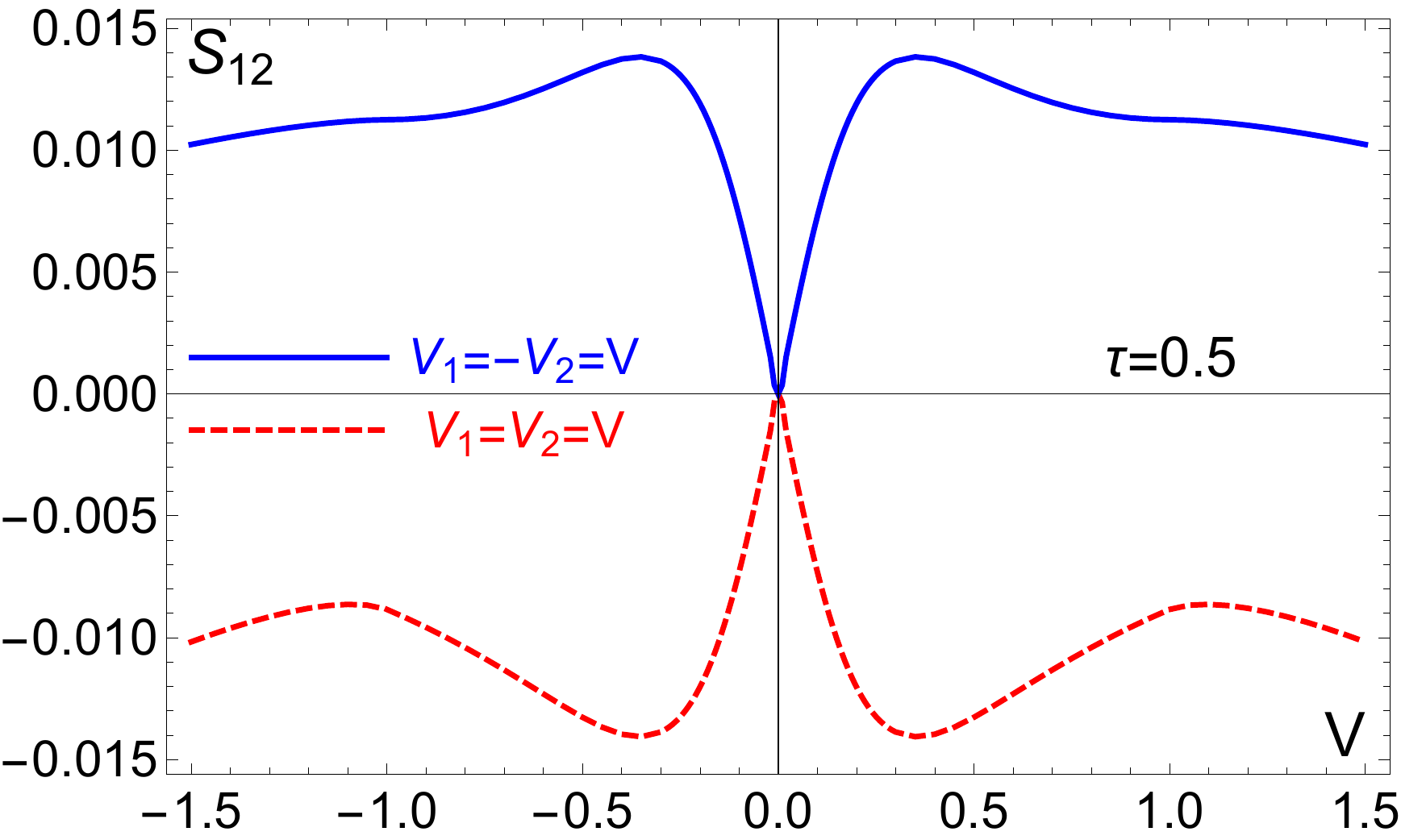} \\
  \includegraphics[width=7.cm]{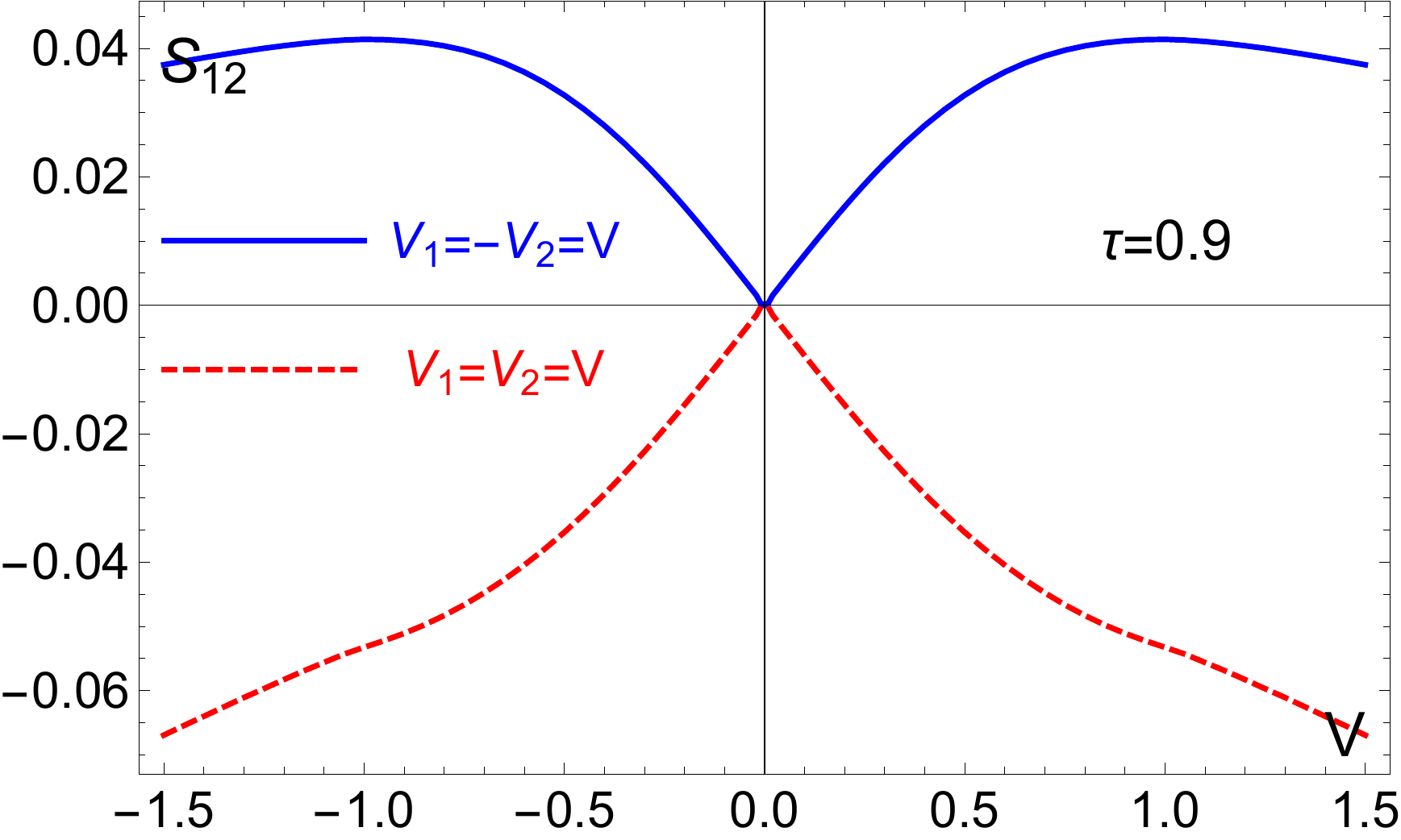} &
 %\centerline{
 \includegraphics[width=7cm]{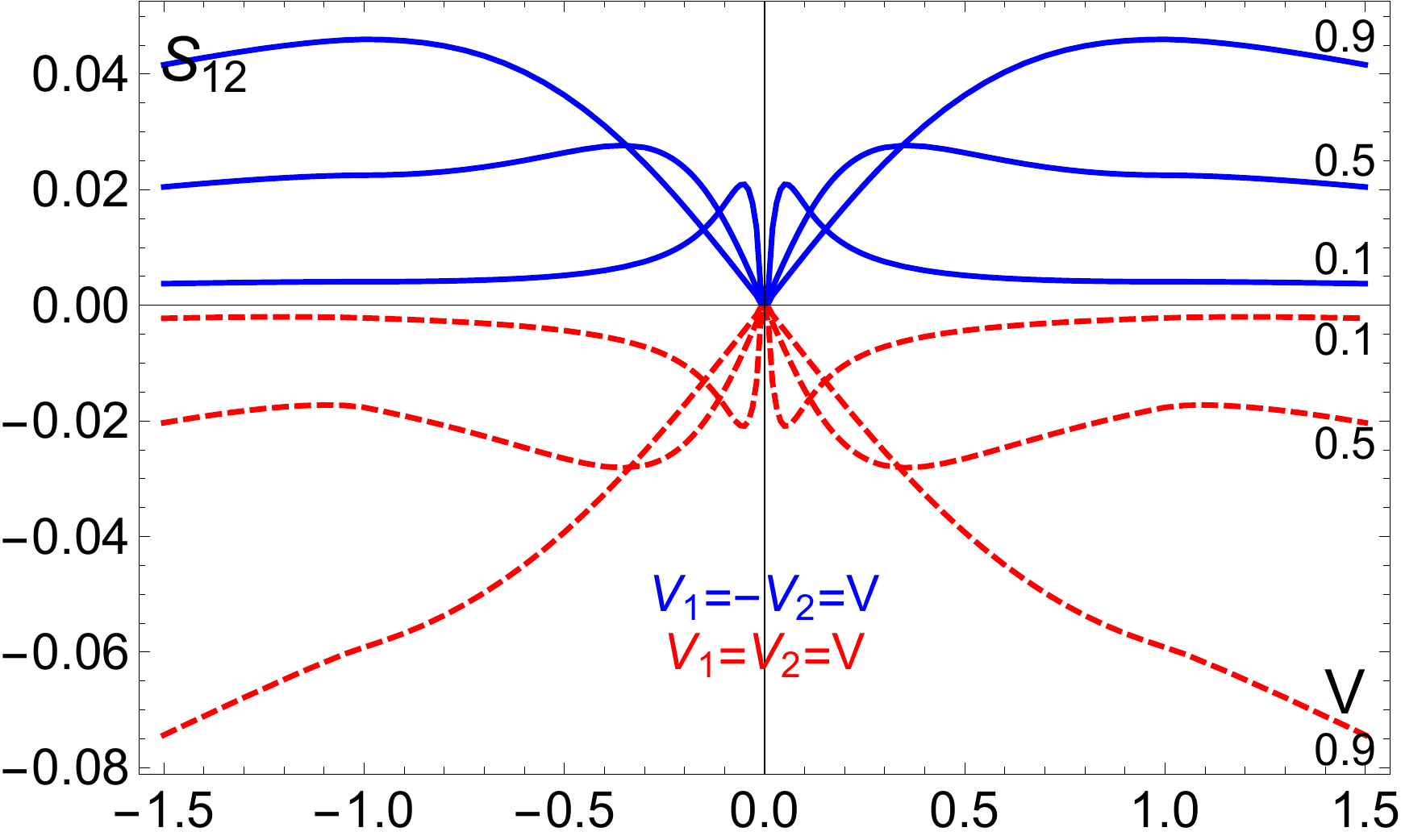} %}
  \end{tabular}
  \caption{HBT cross-correlations $S_{12}$ (in units of $e^2/h$) vs $V=V_1$
   in the case of equal (red dashed curve) and
  opposite (blue full curve) voltages $V_{1,2}$ for $\lambda_1 = \lambda_2$ and several transparencies $\tau$ (as noted on each panel), at
  zero temperature. The bottom right panel combines all curves
  to show the overall scale.
  $S_{12}$ is negative (positive) for equal (opposite) bias voltages, and the values
  of $S_{12}$ are simply opposite in sign in the two cases for $|V| \ll \Gamma$.}
  \label{fig:S12ofV}
\end{figure*}
The expression for the correlations $S_{12}$ in Eq.~\eqref{eq:S12gen} can be further simplified for some specific choices of the bias voltages. For equal voltages $V_1=V_2$, we have
\begin{equation}
 S_{12}(V_1=V_2=V)= -\frac{2 e^2}{h} \frac{\Gamma^2}{4} \frac{|eV|}{(eV)^2+\Gamma^2},
\end{equation}
which coincides with existing results.\cite{haim_current_2015}
 Conversely, for opposite voltages, we have:
\begin{multline}
 S_{12}(V_1=-V_2=V)= \frac{2 e^2}{h} \frac{\Gamma^2}{4} \Big[ \frac{|eV|}{(eV)^2+\Gamma^2}
\\ + \frac{2 \Gamma^2+(eV)^2}{(eV)^2+\Gamma^2} \frac{|eV|}{\Delta^2}
- \frac{2 \Gamma}{ \Delta^2} \mbox{tan}^{-1}\left(\frac{|eV|}{\Gamma}\right)
\Big].
\end{multline}
Fig.~\ref{fig:S12ofV} shows the HBT cross-correlation noise $S_{12}$ for equal
and opposite voltages, computed for three different values of the transmission probability $\tau$. 
One can see that the cross-correlations in these two cases are simply opposite as long as $eV \ll \Gamma$, with negative (positive) values of $S_{12}$ for the equal (opposite) voltage case. 
For $|eV|$ larger than the gap $\Delta$, $S_{12}$ is always a decreasing function
of $|V|$, which, for the opposite voltage case, eventually becomes negative for $|eV| \gg \Delta$ (not shown), as the TS behaves essentially as a normal electrode at such high voltages.

\subsection{Discussion of auto- and crossed- correlations}

In order to better understand the behavior of the HBT noise correlations
$S_{12}$, it is useful to discuss all the noise contributions, in particular the  autocorrelations $S_{00}$, $S_{11}$ and $S_{22}$ are also considered. From
the definition of the noise, Eq.~\eqref{eq:SjjfromG}, and using $-I_0 = I_1 + I_2$, we have
the relation:\cite{martin_wave-packet_1992}
\begin{equation}
S_{00} = S_{11} + S_{22} + 2 \, S_{12}
\label{noises_relation}
\end{equation}
so that is is enough to consider $S_{00}$, $S_{11}$ and $S_{12}$  to achieve a full characterization. For simplicity, we consider in what follows the regime of symmetric couplings ($\lambda_1=\lambda_2$), in the zero-temperature limit.

\subsubsection{Equal voltages}

For equal voltages $V_1=V_2=V$, the analytical expressions for the various noise correlations (in units of $e^2/h$) are
\begin{align}
S_{00} =&  2\Gamma \left[ \mbox{tan}^{-1}\left(\frac{|eV|}{\Gamma} \right) -
          \frac{|eV|/\Gamma}{1+(eV/\Gamma)^2} \right] \simeq 0 \label{eq:S00eqV}\\
S_{11} =&    \Gamma   \mbox{tan}^{-1}\left(\frac{|eV|}{\Gamma} \right) -
         \frac{1}{2}        \frac{|V|}{1+(eV/\Gamma)^2}  \simeq \frac{|eV|}{2}  \label{eq:S11eqV}\\
S_{12} =&       - \frac{1}{2} \frac{|eV|}{1+(eV/\Gamma)^2}  \simeq -\frac{|eV|}{2}  \label{eq:S12eqV}
\end{align}
where the final expressions are obtained in the low-voltage limit $eV \ll \Gamma$. In terms of the total coupling $\Gamma$, the auto-correlation noise $S_{00}$ in the TS lead has the same expression as for a single N-TS junction.\cite{zazunov_low_2016}

\begin{figure*}
 \centerline{\includegraphics[width=12.cm]{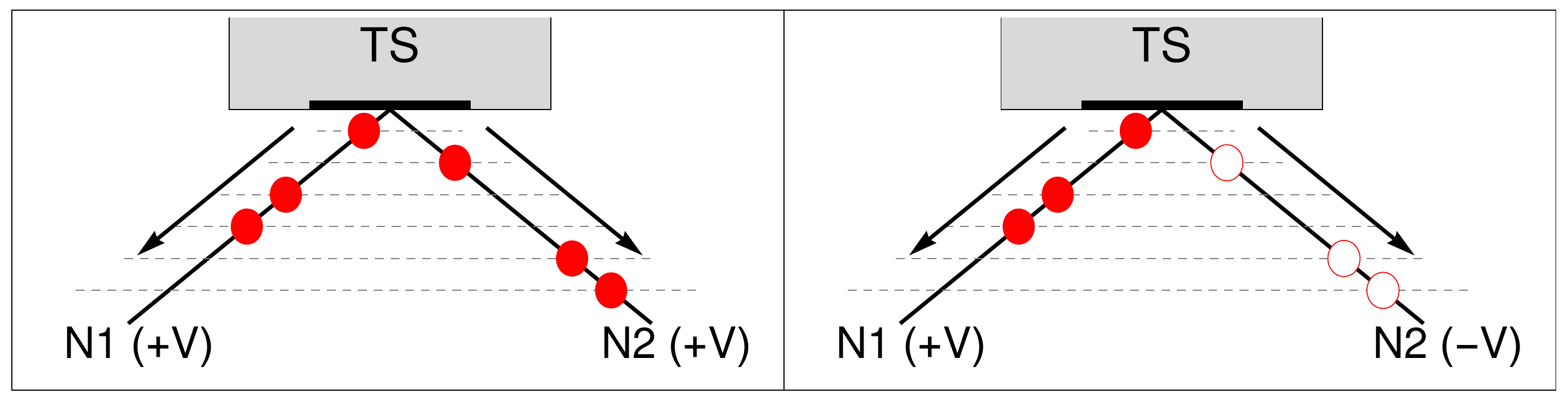}}
 \caption{Schematic picture of the current partitioning between the TS and two normal leads
 (N$_{1,2}$) at low voltage $|eV| \ll \Gamma$. Electrons (holes) are shown as full (empty) circles.
 Left panel: the case of equal voltages, where a noiseless stream of electrons from
 the Majorana bound state is partitioned between the two normal leads, with perfect anti-correlations of the two electron streams.
 Right panel: the case of opposite voltages where lead N$_2$ is biased at potential $-V$, so that electrons are
 emitted into N$_1$ while holes are emitted into N$_2$. The two fermion streams are perfectly anticorrelated,
 which leads to positive cross-correlation noise. The arrows indicate directions of quasiparticle motion.}
 \label{fig:gAntiCorrAll}
\end{figure*}

Focusing on Eqs.~\eqref{eq:S00eqV}-\eqref{eq:S12eqV}, the behavior of the auto- and cross-correlations at low voltage $|eV| \ll \Gamma$ can be understood from the basic properties of the coupling of the normal electrodes to the Majorana bound state.

From the point of view of the TS, the two normal electrodes are at the same potential and thus act as a single one for the total current $I_0$, so that the total conductance has a peak of height $2 e^2/h$ [see Eq.~\eqref{eq:G1} and discussion below]. As a consequence, much like in the single N-TS junction, the total current $I_0$ is noiseless at low voltage $eV\ll \Gamma$, which is confirmed by Eq.~\eqref{eq:S00eqV}.
This total current $I_0= 2(e^2/h) V$  is partitioned here with equal probability between the currents $I_1$ and $I_2$ (see Fig.~\ref{fig:gAntiCorrAll}) :
\begin{equation}
I_1=I_2=(e^2/h)V .
\end{equation} 
These two currents are thus equivalent to the transmitted and backscattered current from a quantum point contact with incoming current $I_0$ and transmission $T=1/2$, reflection $R=1-T=1/2$.

This implies that the autocorrelations $S_{11}$ and $S_{22}$ correspond to the noise associated with currents resulting 
from random partitioning,\cite{blanter_shot_2000} leading to (restoring units)
\begin{equation}
S_{jj} \equiv eI_{j}(1-T) = \frac{e^2}{h} \frac{|eV|}{2} \quad (j=1,2)
\end{equation}
which coincides with Eq.~\eqref{eq:S11eqV} to lowest order in $V$. 

Finally, the HBT noise $S_{12}$ corresponds to the correlation between the two partitioned currents $I_1$ and $I_2$. Due to the fermionic nature of the electrons, these two currents are totally anti-correlated (see Fig.~\ref{fig:gAntiCorrAll}), yielding a negative correlation noise.\cite{buttiker_scattering_1990,martin_wave-packet_1992} 
Following Eq.~\eqref{noises_relation}, and using that $I_0$ is noiseless, one sees that the HBT correlations and the autocorrelations are simply related as $S_{12} = - S_{11}$, which agrees with Eq.~\eqref{eq:S12eqV} to lowest order.

\subsubsection{Opposite voltages}

For opposite voltages $V_1=-V_2=V$ the auto- and cross-correlations take the form (in units of $e^2/h$):
\begin{align}
S_{00} =&  2\Gamma  \mbox{tan}^{-1}\left(\frac{|eV|}{\Gamma} \right) \simeq 2|eV| \label{eq:S00oppV} \\
S_{11} =&    \Gamma   \mbox{tan}^{-1}\left(\frac{|eV|}{\Gamma} \right) -
                \frac{|eV|/2}{1+(eV/\Gamma)^2} - f(V,\Gamma)
            \simeq \frac{|eV|}{2} \label{eq:S11oppV}  \\
S_{12} =&       \frac{1}{2} \frac{|eV|}{1+(eV/\Gamma)^2}   + f(V,\Gamma)
   \simeq \frac{|eV|}{2} \label{eq:S12oppV}
\end{align}
where the final expressions correspond to the low-voltage regime $eV \ll \Gamma$, and we introduced
\begin{equation}
f(V,\Gamma) =  \frac{\Gamma^2}{2 \Delta^2}
  \left[
 eV \, \frac{2 \Gamma^2 + (eV)^2}{(eV)^2+\Gamma^2}
    - 2 \Gamma \mbox{tan}^{-1} \left(\frac{|eV|}{\Gamma}\right)\right]
\end{equation}

While this opposite voltage case has a behavior strikingly different from its equal voltage counterpart, it can still be understood with the same ingredients, by taking into account that the coupling to the Majorana bound state is perfectly electron-hole symmetric. 
Indeed, when normal lead 2 is biased at voltage $-V$ rather
than $V$, it can be seen as a reservoir of holes biased at voltage $V$ coupled to the Majorana bound state.
The behavior of the system is thus the same as for the equal voltage case, except that
electrons are now replaced by holes for the current $I_2$. 
Picturing the total
current from the TS as a stream of \textit{particles} (thus disregarding the charge),
this stream is still noiseless, with one particle (electron or hole)
emitted during each time interval $\hbar/eV$. The currents $I_{1}$ and $I_{2}$ still result from the random partitioning of such a noiseless stream of \textit{particles},
with electrons for $I_1$ and holes for $I_2$, so that
\begin{equation}
I_1=-I_2=(e^2/h) V
\end{equation}
As a consequence the autocorrelation noises
$S_{11}=S_{22}=(e^3/h)|V|/2$ are identical to their counterparts in the equal voltage case, to lowest order in $eV/\Gamma$ [see Eq.~\eqref{eq:S11oppV}].
Much like the equal voltage case, the two currents $I_1$ and $I_2$ are totally anticorrelated, which leads to the same expression for the HBT correlation noise $S_{12}$, only with the opposite sign, as the carriers in the two leads now bear opposite charges (see Fig.~\ref{fig:gAntiCorrAll}). 

Finally the total noise $S_{00}$ (which accounts for the charge of the carriers) corresponds to the current noise of a noiseless stream of \textit{particles} (one particle --electron or hole-- transmitted at each time interval $\hbar/eV$) but with
particles which can be either electron or holes. According to Eq.~\eqref{noises_relation} this creates a total charge noise $S_{00} = 2(e^3/h)|V|$, which coincides with Eq.~\eqref{eq:S00oppV}.

The equal and opposite voltage regimes thus have similar cross-correlation
noises but with opposite signs, a direct consequence of the peculiar properties of the Majorana bound state, which by definition does not distinguish electrons from holes.

\section{Finite length and doping effects} \label{sec:beyond}

The results shown in the previous sections have been obtained for the case
of a semi-infinite topological superconductor bearing a (single) Majorana bound state at its end, with the boundary Green function for the TS nanowire computed in the wide-band limit at half-filling, see Eq.~\eqref{eq:Ginfinite}. 
We wish to address here the case of a finite size TS whose two Majorana bound states (one at each extremity) overlap, leading to radically different behavior for the HBT correlations. Secondly, this TS wire can be doped in such a manner that it becomes a trivial superconductor devoid of topological effects. In this section, we thus briefly discuss how the noise correlations are modified when going beyond the approximations used in the preceding sections.

\subsection{Varying the length of the TS nanowire}

\begin{figure}
\centerline{\includegraphics[width=9.cm]{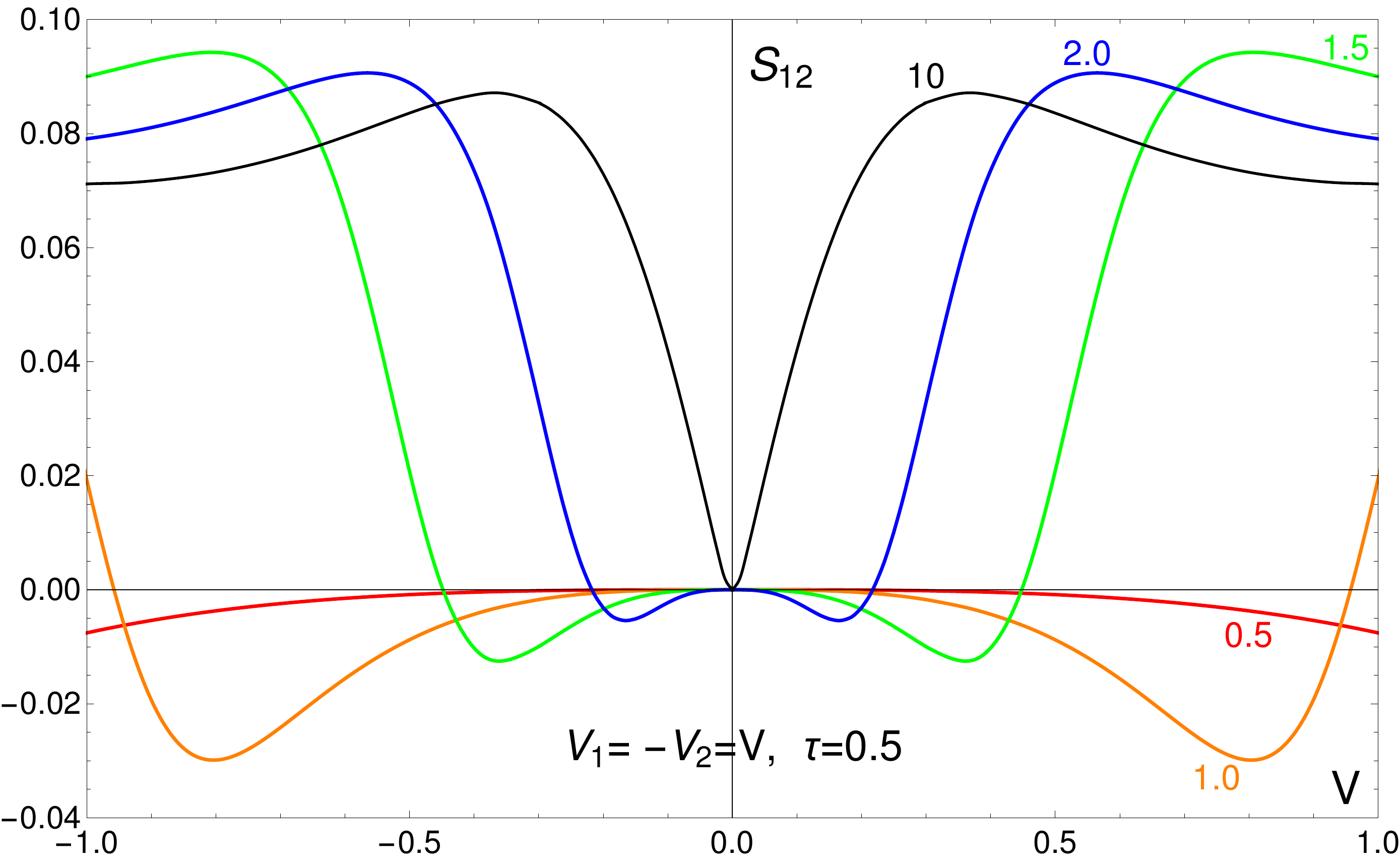}}
\caption{Cross-correlations $S_{12}$ vs $V = V_1$ in the case
of opposite bias voltages ($V_1=-V_2$) for a symmetric junction ($\lambda_1 = \lambda_2$) with total transmission $\tau=0.5$
and several values of the TS wire length $L$. Each curve is labeled by the corresponding value of $L$ in units of $\xi_0=\hbar v_F/\Delta$.
For $L \gg 1$, $S_{12}$ is identical to the infinite TS case,
while for $L \sim 1$ the cross-correlations $S_{12}$ become negative over the range  $|V|/\Delta \lesssim (\xi_0/L)^2$}
\label{fig:S12_diffL}
\end{figure}

The retarded/advanced GF for a TS wire of length $L$ ($-L/2<x<L/2$) computed near $x = L/2$ in the wide-band limit is given by\cite{zazunov_low_2016}
\begin{equation}
g^{R/A}(\omega) = \omega \, \mbox{tanh}(\zeta_{\omega} L)
\frac{\zeta_{\omega} \sigma_0 - \mbox{tanh}(\zeta_{\omega} L ) \Delta \sigma_x}{(\omega \pm i 0^+)^2 - \epsilon_{\omega}^2}
,
\end{equation}
where $\zeta_{\omega} = \sqrt{\Delta^2 - \omega^2}$ and
$\epsilon_{\omega} = \Delta/\mbox{cosh}(\zeta_{\omega}L)$
with $v_F=\hbar=1$.
 The finite TS has a Majorana bound state localized at each end.   When the length
of the TS is much larger than the typical scale of the Majorana bound state,
which is of the order of the superconducting coherence length $\xi_0 = \hbar v_F/ \Delta$,
the two end-state wave functions practically do not overlap and one recovers the same result
as for the infinite length GF. However, with decreasing $L$, this overlap becomes important, and we expect to lose the behavior specific to the presence of a
Majorana bound state. In Fig.~\ref{fig:S12_diffL}, we show the noise correlations $S_{12}$
as a function of the voltage $V$ for the opposite voltage case, at transmission $\tau=0.5$,
and for several values of the TS length $L$ (in units of $\hbar v_F/\Delta$).
  We see that when $L \gg 1$, the results
obtained for $V_1=V=-V_2$ are identical (positive HBT correlations) to the ones obtained in the previous section with the infinite length
Green function.
 However, when $L$ is of order 1, the overlap
of the two Majorana end states becomes important, and $S_{12}$ turns negative around $V=0$, over a range of voltage which increases as $L$ decreases.

\subsection{Varying the chemical potential of the TS nanowire} \label{sec:Smu}

Another important parameter is the intrinsic chemical potential of the topological superconductor,
which depends in a real nanowire on the values of the proximity induced coupling, the
magnetic field, etc. In our approach, this is modeled by the chemical potential $\mu$
of the  Kitaev chain. In the calculations presented so far
we always set $\mu = 0$, corresponding to half-filling of the
chain. 
By varying this parameter at finite bandwidth, it is possible to go away from half-filling and therefore to drive the system from the
topological phase to a trivial one where no Majorana bound state is present. \cite{alice_new_2012}
In order to observe this transition, one needs to rederive the Green function for a Kitaev chain beyond the wide-band limit, with arbitrary values of the gap $\Delta$,
the hopping parameter $t_0$ and the chemical potential $\mu$.  
Explicit formulas for this
Green function, and details of the derivation are provided in App.~\ref{app:Green_mu}.
\begin{figure}
\centerline{\includegraphics[width=9.cm]{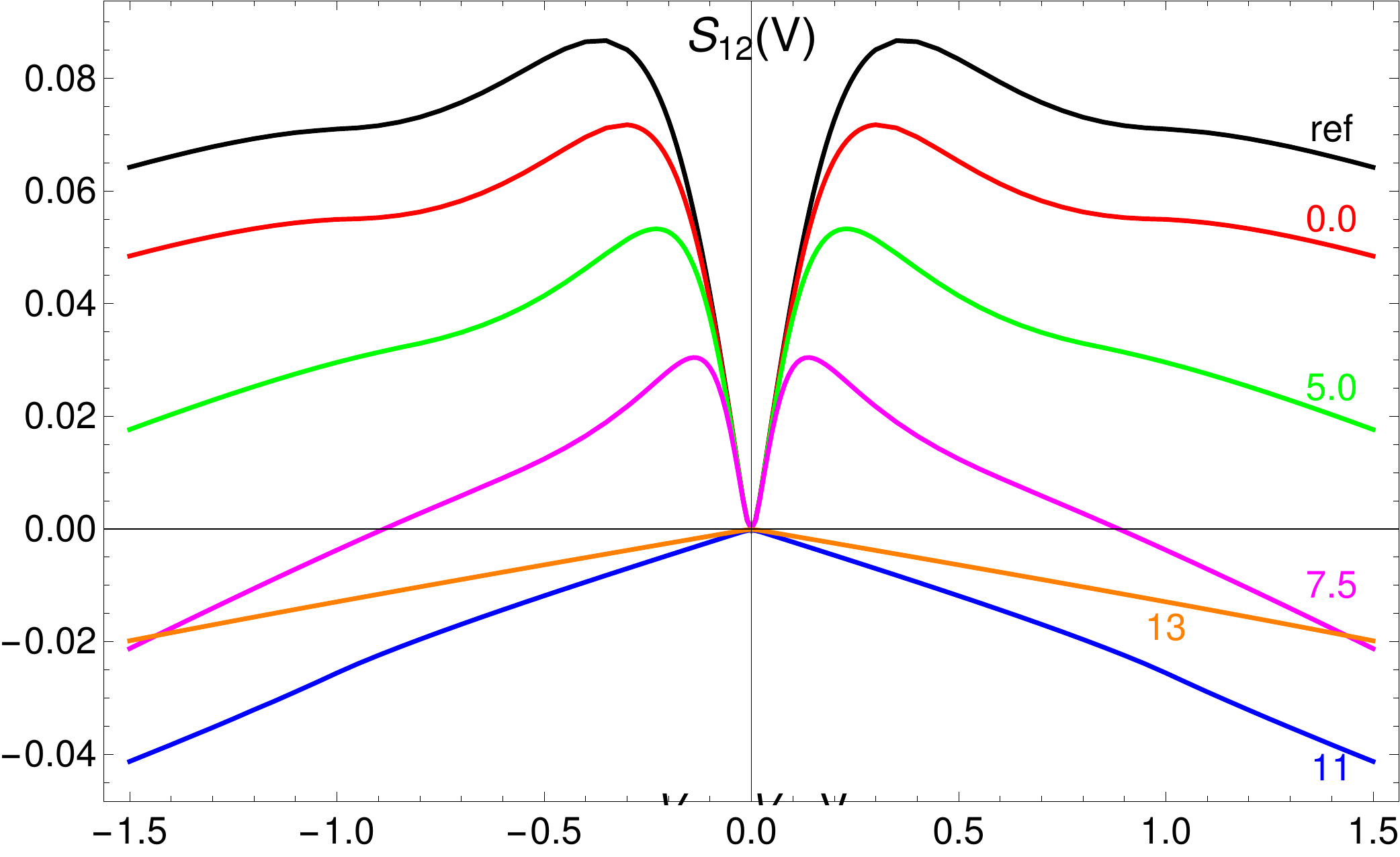}}
\vspace{0.5cm}
\centerline{\includegraphics[width=9.cm]{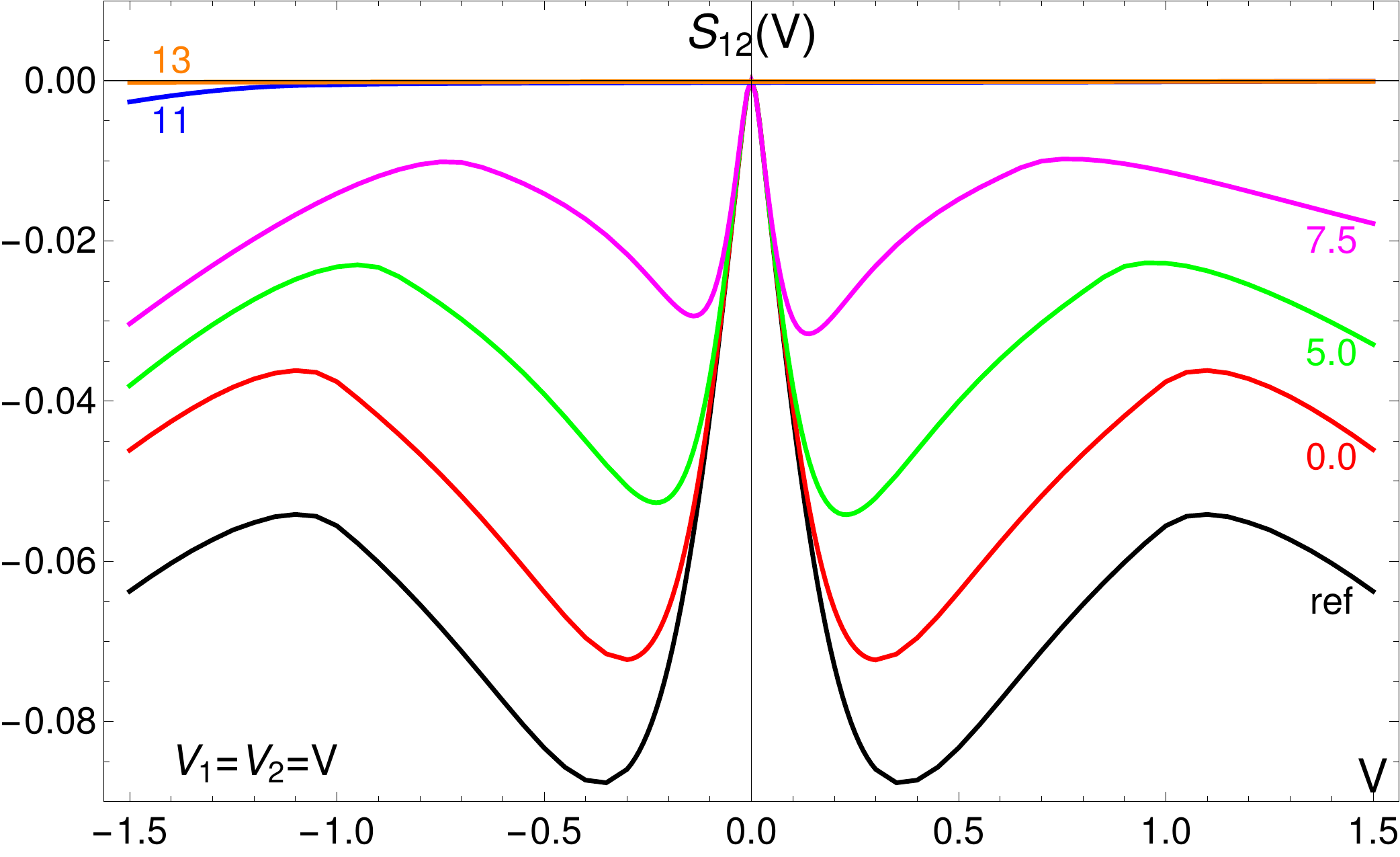}}
\caption{Current correlations $S_{12}$ vs $V=V_1$ in the setup with an infinite TS wire
for several values of its chemical potential $\mu$ (each curve is labeled by the corresponding value of $\mu/\Delta$).
 The top (bottom) panel shows the case of opposite (equal) voltages.
We consider a symmetric junction ($\lambda_1 = \lambda_2$) with a total transparency
$\tau=0.5$, and a hopping strength (TS wire bandwidth) $t_0=10 \Delta$.
The black curve (labeled ``ref'') on each plot corresponds to the wide-band limit [using Eq.~\eqref{eq:Ginfinite}] and serves as a reference for comparison.
There is a clear transition when $\mu$ reaches the bandwidth $t_0$: the peak behavior around $V=0$
disappears when $\mu>t_0$, signaling the absence of a Majorana bound state for $\mu > t_0$.}
\label{fig:S12_mu}
\end{figure}

Fig.~\ref{fig:S12_mu} shows the noise correlation $S_{12}$ for the opposite
(top panel) and equal voltage case (bottom panel),
 for a chain with a hopping parameter
$t_0 = 10 \Delta$, and chemical potential $\mu$ varying from 0 to $13 \Delta$.
When $\mu$ is increased, while still below the bandwidth $t_0$,
the correlations $S_{12}$ are reduced in absolute value, but keep the same qualitative features, with a dip (peak) around $V=0$ for opposite (equal) voltages.
In the opposite voltage case (top panel), $S_{12}$ is always positive for
$V$ close to 0, and the decrease at large voltage becomes more pronounced as $\mu$ is increased. 
In the equal voltage case (bottom panel),
the correlations $S_{12}$ are essentially scaled down when $\mu$ is increased,
with a notable asymmetry between $V>0$ and $V<0$ appearing at large $\mu$.

However, the behavior becomes qualitatively different when $\mu$ reaches the
value of the bandwidth $t_0$, as the peak around $V=0$ disappears for $\mu > t_0$.
 In the opposite voltage case (top panel
of Fig.~\ref{fig:S12_mu}), $S_{12}$ becomes negative for all $V$, even
close to zero voltage. 
This is consistent with our previous interpretation: the positive cross-correlations at low voltage are associated with the coupling to the Majorana bound state
and this feature disappears for $\mu \ge t_0$, when we cross into the trivial non-topological phase.
In the equal voltage case (bottom panel), the specific behavior which was present
for small $V$ also disappears, and the correlations $S_{12}$ become fully asymmetric,
almost vanishing for $V > 0$.
Note that we avoid on purpose to compute the noise $S_{12}$ at 
the precise value of the transition $\mu = t_0$. Indeed the spatial extent of the Majorana bound state  diverges in this case,\cite{kitaev_unpaired_2000} and the
overlap of the Majorana bound states at the two ends
of the system could become important, with a behavior
similar to the one presented in Fig.~\ref{fig:S12_diffL}.

\section{Conclusions}

In this work, we have explored the properties of a topological superconductor nanowire,
including a Majorana bound state at its end, by coupling it to two biased normal leads.
We computed the currents and the Hanbury-Brown and Twiss current cross-correlations,
and showed that the sign of such correlations have a very peculiar dependence on the two voltages $V_1$ and $V_2$ of the normal leads: the correlations are negative when the voltages have the same sign, and become positive when the voltages have opposite signs. 
In addition, for voltages smaller than the coupling between the TS and the normal leads, the correlations for the equal and opposite voltage cases are exactly opposite. This behavior is in stark contrast with the one observed with a conventional BCS superconductor, where  typically correlations are positive for voltages of the same sign only.\cite{buttiker_quantum_2003} This is directly related to the properties of the Majorana bound state, which by definition makes no difference between electrons and holes.
Changing the sign of one of the voltages is equivalent to replacing a
reservoir of electrons with a reservoir of holes - the coupling to the Majorana bound state is unaffected, simply leading to a change of sign of the correlations.

The crossed correlations of a TS wire below the gap at equal voltages 
are similar to the fermionic version
of the HBT experiment in normal metals, with differences showing only at $V > \Gamma$.
%strictly identical to the fermionic version
%of the HBT experiment in normal metals. 
\cite{buttiker_scattering_1990,
martin_wave-packet_1992,buttiker_scattering_1992,henny_fermionic_1999,oliver_hanbury_1999} 
There, the filled Fermi sea which injects electrons at the two outputs is totally noiseless, 
and electron partitioning leads to negative cross-correlations. 
For the TS beam splitter presented here, it is therefore crucial to also probe the HBT noise at opposite voltages to exhibit its positive sign, in order to rule out 
a ``trivial'' interpretation in terms of normal fermionic leads. Positive correlations in this configuration may imply -- as for the BCS Cooper pair splitter\cite{lesovik_electronic_2001} -- that the electron/hole pairs emitted in the two normal metal leads coupled to the TS via the Majorana fermion may form an entangled state.     

Calculations were performed using a Keldysh boundary Green function approach, based
on a Hamiltonian formalism.\cite{zazunov_low_2016}
Each electrode is then represented by its boundary Green function,
and the electrodes are coupled through a tunneling Hamiltonian.
Solving the Dyson equation gives exact (non-perturbative) simple formulas for the current and the current
correlations in terms of the full Green function of the system.
The boundary Green function approach used here is fully equivalent to the scattering matrix approach for voltages below the gap (see App.~\ref{app:Smatrix}), but also allows us to easily access the regime of voltages above the gap.

With this method, we were also able to consider more general situations, simply
by adapting the boundary Green function used for the TS nanowire.
We considered a TS nanowire of finite length and also studied the effect of varying the chemical potential. 
Our results show the existence of a crossover when the length $L$ becomes smaller than $\hbar v_F/\Delta$ (the typical length associated with the Majorana bound state), with positive correlations becoming negative for small length. This behavior is due to the hybridization of the two Majoranas at the ends of the TS nanowire, which then behave as a regular fermion. This confirms that the unusual sign observed for the current correlations for a long (or semi-infinite) TS nanowire is specific to the presence of a Majorana bound state. 
Next, we considered the case where
the TS nanowire is represented by a Kitaev chain with a finite bandwidth, and variable
chemical potential $\mu$ (the corresponding boundary Green
function is derived in App.~\ref{app:Green_mu}).
 By varying $\mu$, one can explore the transition from a topological
superconductor to a conventional one. Our results show that above the transition, the cross-correlations become negative at all voltage, and the specific features due to the Majorana bound state (peak at low voltage) disappear.

We believe that these results, obtained by placing the TS nanowire in a three-lead hybrid system,
give access to unique properties of the TS nanowire, and of Majorana bound states, which
would be more difficult to characterize in a simple two-lead setup. The specific dependence of the
sign of the correlations as a function of the voltages could provide an extremely firm
experimental proof of the presence of a Majorana bound state in a TS nanowire.
Because of its versatility and efficiency, the same
boundary Green function approach could be used to consider other multi-terminal setups involving one or several TS nanowires, or more involved effects, such as interactions.

\acknowledgements

We acknowledge discussions with A. Levy Yeyati.
The authors also acknowledge the support of Grant No. ANR-2014-BLANC (``one shot reloaded'').
This work was granted access to the HPC resources of Aix-Marseille Universit\'e financed by the project Equip@Meso
(Grant No. ANR-10-EQPX-29-01) and has been carried out in the framework of the Labex ARCHIMEDE
(Grant No. ANR-11-LABX-0033) and of the AMIDEX project (Grant No. ANR-11-IDEX-0001-02), all funded by
the ``investissements d'avenir'' French Government program managed by the French National Research Agency (ANR).
This work has also been supported by Deutsche Forschungsgemeinschaft (Bonn) Grant No. EG 96/11-1.
A.~Z. is grateful to the CPT for hospitality during his visit to Marseille.

\appendix

\section{Boundary GF of semi-infinite Kitaev's chain}
\label{app:Green_mu}
In this Appendix, we provide a derivation of the retarded/advanced GF used in Sec.~\ref{sec:Smu},
which describes quasiparticle excitations at the boundary of a semi-infinite TS wire.
The derivation is performed for arbitrary values of the chemical potential $\mu$ and (normal-state) conduction bandwidth $t_0$,
so that by varying $\mu$  one can drive the wire through a topological transition at $\mu = \pm t_0$. \cite{kitaev_unpaired_2000,alice_new_2012}
This generalizes the derivation outlined in Ref.~\onlinecite{zazunov_low_2016} for the case $\mu=0$.
Following the strategy of Ref.~\onlinecite{zazunov_low_2016}, we first compute a ``bulk'' GF for
a homogeneous wire of infinite length, and then the boundary GF is obtained from the Dyson equation of
the wire interrupted by a local potential scatterer.

The TS wire is modeled by a Kitaev chain\cite{kitaev_unpaired_2000,alice_new_2012} representing
an effectively spinless single-channel $p$-wave superconductor.
In terms of fermion operators $c_x$ on a 1D lattice with site numbers $x$ (we set the lattice constant to unity),
the model Hamiltonian reads
\begin{equation}\label{app:HK}
H_K = {1 \over 2} \sum_x \left( -t_0 c^\dagger_x c_{x+1} + \Delta c_x c_{x+1} + {\rm H.c.}\right) - \mu \sum_x c^\dagger_x c_x  ,
\end{equation}
where $t_0 > 0$ is the hopping matrix element, $\Delta > 0$ is the $p$-wave pairing amplitude and $\mu$ is the chemical potential.
Imposing periodic boundary conditions $c_x = c_{x+N}$, with the number of lattice sites $N \rightarrow \infty$,
and passing to momentum space, $c_x = N^{-1/2} \sum_k e^{i k x} \psi_k$,
the ``bulk'' Hamiltonian $H$ takes the standard Bogoliubov-de Gennes form
\begin{equation}
H_K = {1 \over 2} \int_{-\pi}^\pi {dk \over 2 \pi} \, \Psi_k^\dagger h_k \Psi_k ,~~~
h_k = \epsilon_k \sigma_z + \Delta_k \sigma_y ,
\end{equation}
where $\Psi_k = \left( \psi_k, \psi_{-k}^\dagger \right)^T$ is a Nambu spinor
subject to the reality constraint $\Psi_k = \sigma_x \Psi_{-k}^\ast$, $\epsilon_k = - t_0 \cos(k) - \mu$ is the kinetic energy,
$\Delta_k = \Delta \sin(k)$ is the Fourier-transformed pairing,
and Pauli matrices $\sigma_{x,y,z}$ act in Nambu space.
Correspondingly, in coordinate space the retarded/advanced Nambu GF
of $\Psi(x) = \left( c_x, c_x^\dagger \right)^T$ for the Kitaev model \eqref{app:HK} is given by
\begin{equation}
g_{x x'}^{R/A}(\omega) = \int_{-\pi}^\pi {dk \over 2 \pi} \, e^{i k (x - x')}  \left( \omega - h_k \right)^{-1} ,
\end{equation}
where the frequency $\omega$ should be understood as $\omega +i0^+$ ($\omega - i0^+$) for the retarded (advanced) GF.
Introducing a new integration variable $z = - \cos(k)$, some algebra yields:
\begin{multline}\label{app:g}
g_{x x'}^{R/A}(\omega) = {1 \over 2 \pi (\Delta^2 - t_0^2)} \int_{-1}^1 {dz \over \sqrt{1 - z^2} D(z, \omega)} \\
\sum_{s = \pm} \left[ \omega + (t_0 z - \mu) \sigma_z + s \Delta \sqrt{1 -z^2} \sigma_y \right] \\ \times \left( -z + i s \sqrt{1 -z^2} \right)^{x-x'} ,
\end{multline}
with
\begin{align}
D(z, \omega) &= (z - {\cal Q}_+) (z - {\cal Q}_-)  \nonumber \\
{\cal Q}_\pm(\omega) &= \frac{-\mu t_0 \pm \sqrt{\mu^2 t_0^2 - (\Delta^2 - t_0^2) (\omega^2 - \Delta^2 - \mu^2)}}{\Delta^2 - t_0^2} .
\end{align}

The boundary GF of a semi-infinite Kitaev chain located at $x>0$
can be obtained from the ``bulk'' GF for the translationally invariant model \eqref{app:HK}
by adding a local impurity of strength $U$ at site $x=0$,
which results in the Hamiltonian $\tilde H_K = H_K + U c^\dagger_0 c_0$.
The ``full'' retarded/advanced GF $\tilde g_{x x'}^{R/A}(\omega)$ then obeys the following Dyson equation:
\begin{equation}\label{app:Dysoneq}
\tilde g_{x x'}^{\nu = R/A}(\omega) = g_{x x'}^\nu(\omega) + g_{x 0}^\nu(\omega) U \sigma_z \tilde g_{0 x'}^\nu(\omega) .
\end{equation}
In the limit $U \rightarrow \infty$, i.e., when one effectively cuts the wire into two semi-infinite pieces,
Eq.~\eqref{app:Dysoneq} yields for the boundary GF\cite{zazunov_low_2016} defined as $G(\omega) = \tilde g_{1 1}(\omega)$
\begin{equation}\label{app:G}
G^{\nu = R/A}(\omega) = g^\nu_{00}(\omega) - g^\nu_{10}(\omega) \left[ g^\nu_{00}(\omega) \right]^{-1} g^\nu_{01}(\omega) .
\end{equation}
Thus, for computing the boundary GF \eqref{app:G} one only needs to evaluate Eq.~\eqref{app:g} for $x = x'$ and
nearest-neighbor sites $x - x' = \pm 1$.
Using the following integrals with a complex-valued parameter $a$, ${\rm Im} \, a \neq 0$,
\begin{equation}
{1 \over \pi} \int_{-1}^1 {dz \over \sqrt{1 - z^2} (z - a)} = {-1/a \over \sqrt{1 - 1/a^2}} ,
\end{equation}
and for $n=1,2$
\begin{equation}
{1 \over \pi} \int_{-1}^1 {dz \, z^n \over \sqrt{1 - z^2} (z - a)} = a^{n-1} \left( 1 - {1 \over \sqrt{1 - 1/a^2}} \right) ,
\end{equation}
after some algebra we obtain from Eq.~\eqref{app:g}:
\begin{align}
g_{00}^{\nu = R/A}(\omega) &= (\omega - \mu
\sigma_z) {\cal F}_{-1}(\omega)  +  t_0 \sigma_z {\cal F}_0(\omega) , \nonumber \\
g_{\pm 1, 0}^\nu(\omega) = g_{0, \mp 1}^\nu(\omega) &=
\pm i \Delta  {\cal F}_{-1}(\omega) \sigma_y  - (\omega - \mu \sigma_z) {\cal F}_0(\omega)
\nonumber \\
& \quad + \left( t_0 \sigma_z \pm i \Delta \sigma_y \right) \left[ 1 - {\cal F}_1(\omega) \right],
\label{app:g00}
\end{align}
where
\begin{multline}
{\cal F}_{m = 0, \pm 1}(\omega) = {1 \over (t_0^2 - \Delta^2) ({\cal Q}_+(\omega) - {\cal Q}_-(\omega))} \\
\times \sum_{s = \pm} \frac{s {\cal Q}_s^m(\omega)}{\sqrt{1 - 1/{\cal Q}_s^2(\omega)}} .
\end{multline}
The boundary GF of the semi-infinite Kitaev chain then follows by inserting Eq.~\eqref{app:g00} into Eq.~\eqref{app:G}.
Eq.~\eqref{app:g00} is an extension of the result of Ref.~\onlinecite{zazunov_low_2016} to the general case
of $\mu \neq 0$. In particular, for $\mu = 0$ and assuming the wide-band limit $t_0 \gg \max (\Delta, |\omega|)$
the above expressions in Eq.~\eqref{app:g00} simplify to
\begin{align}
g_{00}^{\nu = R/A}(\omega) &= {\omega \over t_0 \sqrt{\Delta^2 - \omega^2}} \, \sigma_0 ,  \nonumber \\
g_{\pm 1,0}^\nu(\omega) &= {1 \over t_0 \sqrt{\Delta^2 - \omega^2}} \, \left(
\sqrt{\Delta^2 - \omega^2} \sigma_z \mp i \Delta \sigma_y \right) ,
\end{align}
and then using Eq.~\eqref{app:G} one recovers the boundary GF \eqref{eq:Ginfinite} quoted in Sec.~\ref{sec:model}.

\section{Analytical formulas for the current and noise}
\label{app:formulas}
This Appendix contains more general formulas for the currents and 
current correlations, which were too lengthy to be shown in the main text.

The expressions \eqref{eq:J11_J12} for the local and non-local differential conductances can be extended for energies above the gap,
leading to the following general forms
\begin{widetext}
\begin{align}
 J_{11}(\omega) &= \left\{
   \begin{array}{lc} \frac{ 2 \lambda_1^2 \lambda_2^2 \left(\Lambda^4-1\right) \frac{\omega^2}{\Delta^2}+ 4 \lambda_1^2
   \Lambda^2} {\left(1-\Lambda^4\right)^2 \frac{\omega^2}{\Delta^2}+ 4\Lambda^4}
                                   & \quad |\omega|<\Delta\\
     \frac{-2\lambda_1^4 \left(\Lambda^4+2 \Lambda^2 \sqrt{1-\frac{\Delta ^2}{\omega^2}}-
     \frac{2 \Delta^2}{\omega^2}+1\right)+2\lambda_1^2 \left(\left(3 \Lambda^4+1\right) \sqrt{1-\frac{\Delta^2}{\omega^2}}+\Lambda^2
     \left(\Lambda^4+3\right)-\frac{2 \Delta^2
   \Lambda^2}{\omega^2}\right)}{\left(\Lambda^4+2 \Lambda^2 \sqrt{1-\frac{\Delta^2}
{\omega^2}}+1\right)^2}
     & \quad |\omega| >\Delta
     \end{array} \right. \\
     J_{12}(\omega) &= \left\{
   \begin{array}{lc}
     \frac{-2 \lambda_1^2 \lambda_2^2 \left(\Lambda^4-1\right) \frac{\omega^2}{\Delta^2}} {\left(1-\Lambda^4\right)^2
     \frac{\omega^2}{\Delta^2}+4 \Lambda^4}  & \quad |\omega|<\Delta\\
     \frac{-2 \lambda_1^2 \lambda_2^2 \left(\Lambda^4+2 \Lambda^2 \sqrt{1-\frac{\Delta^2}{\omega^2}}-
     \frac{2 \Delta^2}{\omega^2}+1\right)}{\left(\Lambda^4+2 \Lambda^2
   \sqrt{1-\frac{\Delta^2}{\omega^2}}+1\right)^2}      & \quad |\omega| >\Delta
      \end{array} \right.
\end{align}
%\end{widetext}

Similarly, one can obtain closed-form expressions for the noise cross-correlations $S_{12}$ at zero temperature for generic
values of the coupling constants $\lambda_1, \lambda_2$ and voltages $V_1, V_2$, thus generalizing Eq.~\eqref{eq:S12gen}.
In the subgap regime $|V_1|, |V_2| \leq \Delta$, this reads
\begin{align}
S_{12} (V_1, V_2) = \frac{e^2}{h} \frac{4 \Delta \left( \lambda_1 \lambda_2 \right)^2}{\left( \Lambda^4 - 1\right)^3}
&
\left\{
- \frac{\left( \Lambda^4 -1 \right)^2}{\Lambda^2} \frac{\Gamma |e V_2|}{\Gamma^2 + (e V_2)^2} \text{sgn} (V_1 V_2) +
2 \lambda_1^2 \left( \Lambda^2 \delta - 1 \right) \left[2 \, \mbox{tan}^{-1} \left( \frac{|e V_1|}{\Gamma} \right) -
\frac{\Gamma |e V_1|}{\Gamma^2 + (e V_1)^2} \right]
\right. \nonumber \\
&
- 2 \lambda_2^2 \left( \Lambda^2 \delta + 1 \right) \left[\mbox{tan}^{-1} \left( \frac{|e V_1|}{\Gamma} \right) +\mbox{tan}^{-1} \left(
\frac{|e V_2|}{\Gamma} \right) - \frac{\Gamma |e V_2|}{\Gamma^2 + (e V_2)^2} \right] \nonumber \\
&
\left.
+ 2 \Lambda^2 \left(1-\delta^2 \right) \left[ \frac{|e V_1|}{\Gamma} + \left( 2 \, \mbox{tan}^{-1} \left( \frac{|e V_2|}{\Gamma} \right) -
\frac{\Gamma |e V_2|}{\Gamma^2 + (e V_2)^2} -  \frac{|e V_2|}{\Gamma} \right) \text{sgn} (V_1 V_2) \right] \right\}
\end{align}
where we introduced $\delta = \lambda_1^2 - \lambda_2^2$ and focused on voltages $|V_1| \geq |V_2|$.
The complementary situation $|V_1| < |V_2|$ is readily obtained from the above expression upon exchanging
$V_1 \leftrightarrow V_2$, $\lambda_1 \leftrightarrow \lambda_2$ (which also implies changing $\delta$ into $-\delta$).
\end{widetext}

\section{Discussion of the case $\Lambda >1$ vs. $\Lambda<1$}
\label{app:Smatrix}

Our microscopic model contains the two parameters $\lambda_1,\lambda_2$ describing the tunneling
from the TS to normal electrodes 1 and 2. As shown in the formulas
for the currents and current correlations,  a natural parameter is
\begin{equation}
\Lambda = \sqrt{\lambda_1^2 + \lambda_2^2}
\end{equation}
which is related to the total transmission probability $\tau$ between the TS and the normal electrodes
\begin{equation}
\tau= \frac{ 4 \Lambda^2}{(1+\Lambda^2)^2}
\label{eq:tauLambda}
\end{equation}
This relation generalizes the equivalent one for a simple junction composed of two
leads.\cite{cuevas_hamiltonian_1996,zazunov_low_2016}
We see that when $\Lambda$ goes from 0 to 1, $\tau$ also varies from 0 to 1, so the whole range
of transparencies is covered by taking $\Lambda$ in $[0,1]$.
From Eq.~\eqref{eq:tauLambda}, one also sees that for a given
$\Lambda$ in $[0,1]$, the value $1/\Lambda$ gives the same value of the transparency $\tau$.
However, as we show below, taking $\Lambda >1$ leads to a different physical realization of the
system.

\begin{figure}
\centerline{\includegraphics[width=6.cm]{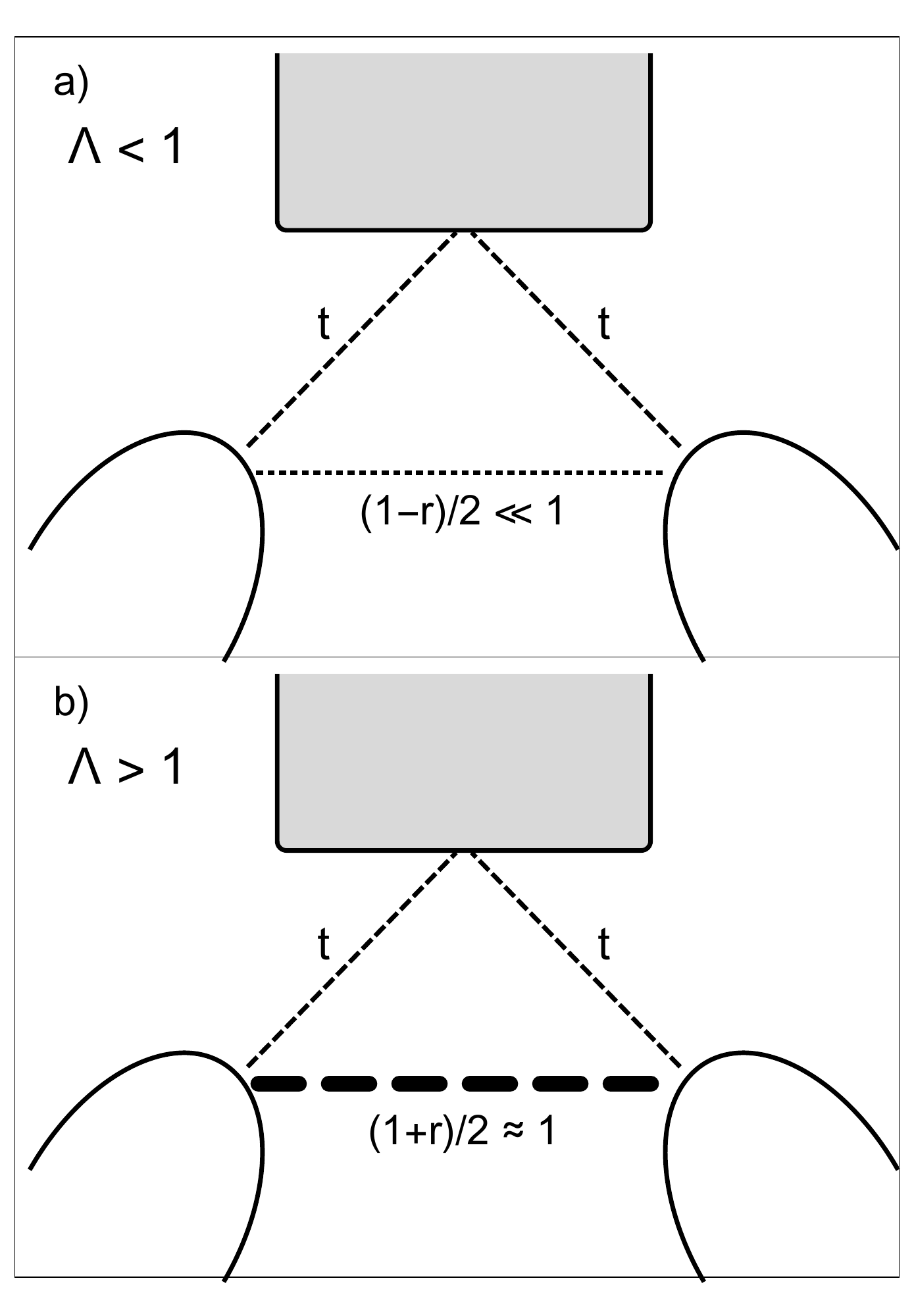}}
\centerline{\includegraphics[width=6.cm]{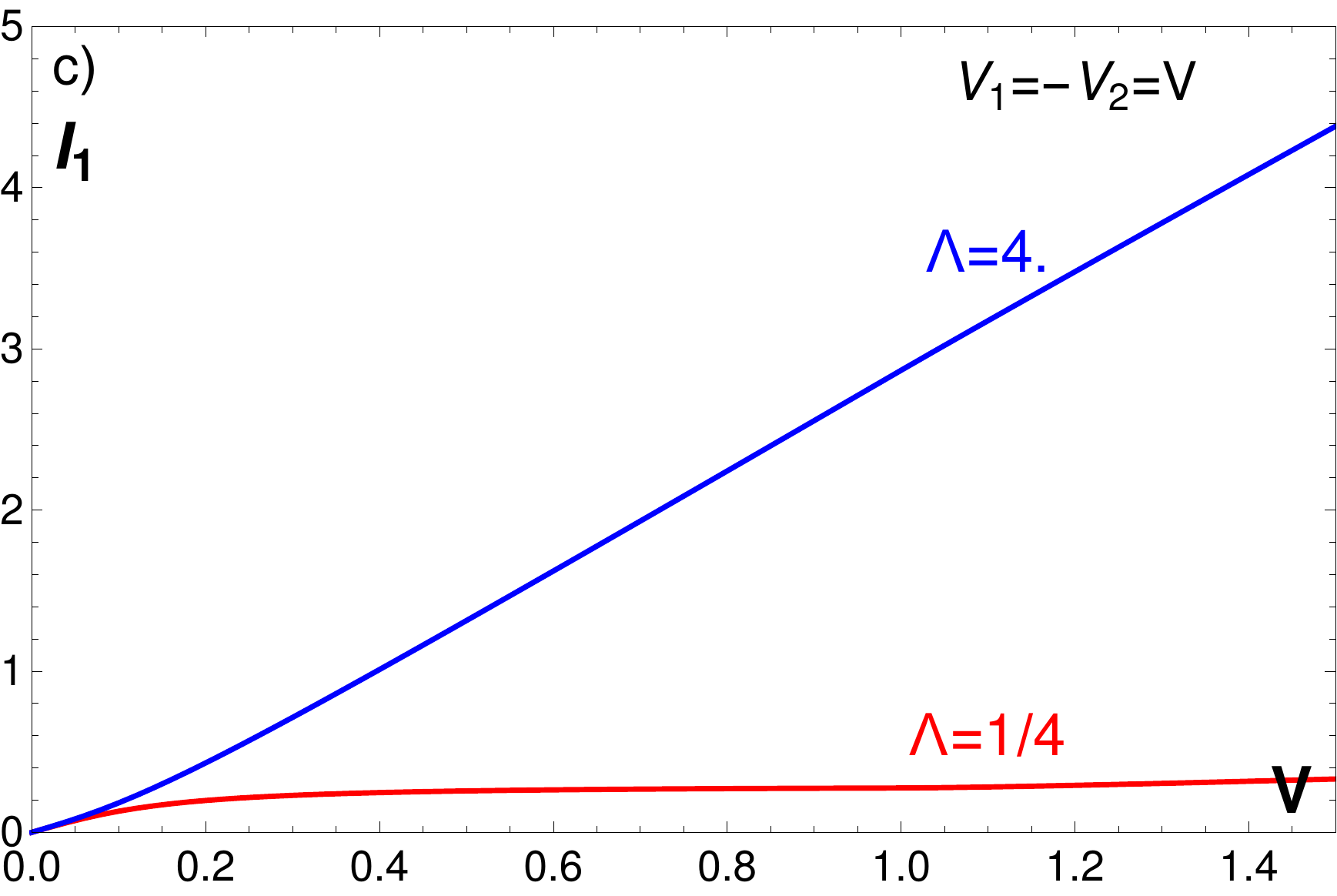}}
\caption{Panels (a) and (b) : schematic illustration of the behavior of the system for the choice of $\Lambda<1$  and $\Lambda>1$ ,
in the case of a strong reflection ($r$ close to 1).
While $\Lambda=\Lambda_0<1$ and $\Lambda=1/\Lambda_0>1$
correspond to the same transparency between the TS and
the normal leads, the transparency of the direct channel between
the two normal leads is totally different in the two cases.
Panel (c) : plots of the current $I_1(V)$ in the $V_1=-V_2=V$ configuration, for $\Lambda=4$ and
$\Lambda=1/4$.}
\label{fig:graph_S12pm}
\end{figure}

Indeed, choosing the value $\tau$ of the total transparency between the TS and the normal leads,
 even for a symmetric system, does not totally specify the system. This can be understood simply
 from the scattering matrix formalism, as noted by Valentini
 and co-workers~\cite{valentini_finite_2016}.
 Writing the scattering matrix for a three-lead system, where the two lateral leads
 (lines/columns 1-2) are symmetrically connected to a central one (line/column 3) with
 a transparency $\tau = \sqrt{1-r^2}$, we have
 \begin{equation}
S = \left(
\begin{array}{ccc}
 \cdot & \cdot & t \\
 \cdot & \cdot & t \\
 t  & t & r
\end{array}   \right)
\label{eq:Smatgen}
\end{equation}
with $r^2 + 2 t^2 =1$.
Imposing the unitarity of the $S$ matrix gives the values of the coefficients
in the 1-2 block $s_{12}$ [written as dots in Eq.~\eqref{eq:Smatgen}],
which represent direct reflection/transmission
in the subsystem of the two lateral leads. There are
two possible solutions (written here for simplicity with real coefficients)
 \begin{equation}
s_{12,+} = \frac{1}{2} \left(
\begin{array}{cc}
 (1-r) & -(1+r)  \\
 -(1+r) & (1-r)
\end{array}   \right)
\label{eq:s12p}
\end{equation}
and
 \begin{equation}
 s_{12,-} = \frac{1}{2} \left(
\begin{array}{cc}
 -(1+r) & (1-r)  \\
 (1-r) & -(1+r)
\end{array}   \right)
\label{eq:s12m}
\end{equation}
and we note $S_{+}$ and $S_{-}$ the complete scattering matrix corresponding to the choice of
$s_{12,+}$ and $s_{12,-}$ respectively.
The difference between the two choices can be understood for example by taking $r$ close to 1 (very poor transmission to the central lead).
Then the amplitude of
direct transmission between 1 and 2 [elements (1,2) and (2,1)]
are very different for $S_{+}$ and $S_{-}$ :
for $S_{-}$, it is $(1-r)/2$, which is close to 0; while for
$S_{+}$ it is $-(1+r)/2$ which is close to one in absolute value.
This means that for a given transmission to the central lead (which fixes $r$),
the currents $I_1$ and $I_2$ have totally different values for the two choices
  $S_{+}$ or $S_{-}$ as soon as the voltages
$V_1$ and $V_2$ are not equal (if $V_1=V_2$, then the direct transmission
 between leads 1 and 2 has no consequence).  The impact of the choice of $S_{+}$ or $S_{-}$ is illustrated in Fig.~\ref{fig:graph_S12pm},
 with a plot showing the current $I_1(V)$ in the opposite
 voltage configuration ($V_1=-V_2=V$), for $\Lambda=1/4$ and $\Lambda=4$. The two
 values of $\Lambda$ correspond to the same transparency $\tau \simeq 0.22$
  between the TS and the two normal  leads, but the currents $I_1$ and $I_2=-I_1$
   are very different for the two values of $\Lambda$.
 For $\Lambda =4$ the current $I_1$ is much larger as $V$ increases, because of the direct current
 going from normal lead 1 to normal lead 2.

The two choices for $S$ correspond precisely to the choice $\Lambda<1$ ($\to S_{-}$) or $\Lambda>1$ ($\to S_{+}$) in the microscopic calculation.
The relation between $\Lambda$ and $r$ is:
\begin{equation}
 \Lambda = \sqrt{\frac{1-r}{1+r}} \quad (\Lambda <1) \quad \quad \mbox{or} \quad \quad
     \Lambda = \sqrt{\frac{1+r}{1-r}} \quad (\Lambda >1)
\end{equation}
or equivalently
\begin{equation}
r = \frac{1-\Lambda^2}{1+\Lambda^2}\quad (\Lambda <1) \quad \quad \mbox{or} \quad \quad
  r = \frac{\Lambda^2-1}{\Lambda^2+1}\quad (\Lambda >1)
\end{equation}
%%%%%%%%%%%%%%%%%%%%%%%%%%%%%%%%%%%%%%%%
with $0<r<1$, and $\Lambda^2 = \lambda_1^2 + \lambda_2^2$ is given from the Hamiltonian.
 As a proof of these relations, 
 we show below that the currents obtained from the scattering matrices
$S_{+}$ and $S_{-}$, for voltages smaller than the gap, coincide with the expressions obtained
from our Hamiltonian calculation in terms of $\Lambda$ [see Eq.~\eqref{eq:J11_J12}].
The details of the scattering matrix calculation follows Ref.~\onlinecite{haim_current_2015}.

Channels 1 and 2 correspond to the two normal leads, while channel 3 is related to the
superconductor. We first construct the $s^{ee}$ and $s^{he}$ 2x2 matrices describing the scattering between the normal leads, of an electron into an electron (ee) or a hole (he).
 These are~\cite{haim_current_2015} :
\begin{align}
s^{ee} & = s_{12} - \frac{a(\omega)^2 r}{1+ r^2 a(\omega)^2}  
 \left( \begin{array}{cc} t^2 & t^2 \\ t^2 & t^2 \end{array} \right) 
 \label{eq:see}  \\
s^{he} & = \frac{a(\omega)^2}{1+ r^2 a(\omega)^2}  
 \left( \begin{array}{cc} t^2 & t^2 \\ t^2 & t^2 \end{array} \right)
 \label{eq:she}  
\end{align}
where $a(\omega) = \mbox{exp}[-i \, \mbox{arccos}(\omega/\Delta)]$ is the amplitude for 
Andreev reflection at energy $\omega$, $s_{12}$ is given by Eq.~\eqref{eq:s12p} or \eqref{eq:s12m},
and $r=\sqrt{1-2 t^2}$ is the reflection amplitude from the superconductor.  For
simplicity, we consider a symmetric system, and we take $r$ and $t$ real.

The expression for the current $I_1$ in terms of the scattering matrix elements is\cite{haim_current_2015}
\begin{align}
I_1 = \frac{e}{h} \int_0^{\infty} \!\!\!\! d\omega \; 
 & n_{F,1e}(\omega)  \left( 1 - \left| s^{ee}_{11} \right|^2 +  \left| s^{he}_{11} \right|^2 \right)
             \nonumber \\
 -& n_{F,1h}(\omega)  \left( 1 + \left| s^{eh}_{11} \right|^2 -\left| s^{hh}_{11} \right|^2 \right)
             \nonumber   \\
 +& n_{F,2e}(\omega)  \left(- \left| s^{ee}_{12} \right|^2 +  \left| s^{he}_{12} \right|^2 \right)
             \nonumber \\    
 +& n_{F,2h}(\omega)  \left( - \left| s^{eh}_{12} \right|^2 +\left| s^{hh}_{12} \right|^2 \right)                                
\end{align}
where $n_{F,1e}(\omega)$ is the Fermi function of electrode 1, with 
$n_{F,1h}(\omega) = 1 -n_{F,1e}(-\omega)$, and $s^{hh}(\omega) = [s^{ee}(-\omega)]^*$,
$s^{eh}(\omega) = [s^{he}(-\omega)]^*$. Comparing with Eq.~\eqref{eq:Igeneral}, we see that
the two expressions of the current are identical if we have $1 - \left| s^{ee}_{11} \right|^2 +  \left| s^{he}_{11} \right|^2 = 2 J_{11}(\omega)$, and $- \left| s^{ee}_{12} \right|^2 +  \left| s^{he}_{12} \right|^2=2 J_{12}(\omega)$.
 
 Using Eqs.~\eqref{eq:see}-\eqref{eq:she}, we get after some algebra 
 (recalling that $r^2+ 2 t^2=1$):
 \begin{align}
  1 - \left| s^{ee}_{11} \right|^2 +  \left| s^{he}_{11} \right|^2 &=
   \frac{\left(1-r^2\right)^2 \pm (1-r)^2 r \omega ^2}{4 r^2 \omega ^2+\left(1-r^2\right)^2} \\
     \left(- \left| s^{ee}_{12} \right|^2 +  \left| s^{he}_{12} \right|^2 \right) &=
    \mp \frac{r (r+1)^2 \omega^2}{4 r^2 \omega^2+\left(1-r^2\right)^2}
 \end{align}
where the  $\pm$ sign refers to the choice of $(s_{12,+})$ or $(s_{12,-})$.  
One can show that these expressions are equal to 2 $J_{11}(\omega)$ and
2 $J_{12}(\omega)$ from Eq.~\eqref{eq:J11_J12}
with $\lambda_1=\lambda_2=\Lambda/\sqrt{2}$ 
if the relation between $r$ and $\Lambda$ is
\begin{equation}
r = \frac{1-\Lambda^2}{1+\Lambda^2}\quad (\mbox{for } s_{12,-}) \quad  \mbox{or} \quad
  r = \frac{\Lambda^2-1}{\Lambda^2+1}\quad (\mbox{for }  s_{12,+})
\end{equation}
%%%%%%%%%%%%%%%%%%%%%%%%%%%%%%%%%%%%%%%%
For a given transparency $\tau$, taking $\Lambda >1$ thus represents a system where there is
a strong, direct link between the lateral leads 1 and 2, which is not the system we
are studying here. This explains why, in all the
results presented in this work, we consider $0<\Lambda<1$ only.

\bibliography{/home/jonckhee/cpt/biblio/BibAll}{}

\end{document}